\documentclass[12pt]{article}
\usepackage[dvips]{graphicx} 
\begin{document}

\begin{center} {\large \bf
Combined Effects of Small Scale Anisotropy and Compressibility
on Anomalous Scaling of a Passive Scalar}\bigskip\\
\end{center}

\begin{center}  {\large  E. Jur\v{c}i\v{s}inov\'a$^{1,2}$, M.
Jur\v{c}i\v{s}in$^{1,3}$, R.\,Remecky$^{4}$, and M.\,Scholtz$^{4}$}
\vspace*{1.5mm}
\end{center}

\begin{center}
{\it $^1$ Institute of Experimental Physics, Slovak Academy of
  Sciences, \\ Watsonova 47, 04001 Ko\v{s}ice, Slovakia} \\
{\it $^2$ Laboratory of Information Technologies, Joint Institute
for Nuclear Research,\\ 141 980 Dubna, Moscow Region, Russian Federation} \\
 {\it $^3$ Laboratory of Theoretical Physics, Joint Institute for
  Nuclear Research, \\ 141 980 Dubna, Moscow Region, Russian Federation
  } \\
{\it $^3$ Department of Physics and Astrophysics, Institute of
Physics,
\\ P.J. \v{S}af\'arik University, Park Angelinum 9, 04001 Ko\v{s}ice, Slovakia}

 e-mail: jurcisin@thsun1.jinr.ru\footnote{Corresponding author: M. Jurcisin,
Laboratory of Theoretical Physics, Joint Institute for Nuclear
Research, 141 980 Dubna, Moscow Region, Russian Federation, tel.: 7
49621 63389; fax.: 7 49621 65084}, eva.jurcisinova@post.sk, \\
richard.remecky@student.upjs.sk, scholtzzz@pobox.sk
\end{center}

\begin{abstract}
Model of a passive scalar field advected by the compressible
Gaussian strongly ani\-so\-tro\-pic velocity field with the
covariance $\propto \delta(t-t^{\prime})|{\bf x}-{\bf
x^{\prime}}|^{2\varepsilon}$ is studied by using the field theoretic
renormalization group and the operator product expansion. The
inertial-range stability of the corresponding scaling regime is
established. The anomalous scaling of the single-time structure
functions is studied and the corresponding anomalous exponents are
calculated. Their dependence on the compressibility parameter and
anisotropy parameters is analyzed. It is shown that, as in the
isotropic case, the presence of compressibility leads to the
decreasing of the critical dimensions of the important composite
operators, i.e., the anomalous scaling is more pronounced in the
compressible systems. All calculations are done to the first order
in $\varepsilon$.
\end{abstract}



\section{\label{sec:level1}Introduction}

During the last two decades the so-called "rapid change model" of a
passively advected scalar by a self-similar Gaussian
$\delta-$correlated in time velocity field introduced by Kraichnan
\cite{Kra68} and number of its extensions have played the central
role in the theoretical investigation of intermittency and anomalous
scaling, the problems which stay in the center of attention in the
framework of the inertial range investigation of fully developed
turbulence \cite{MonYagBook,HuPhWi91}, i.e., in the range
characterized by the scales which are far away from the largest
scales at which energy is pumping into the system and, at the same
time, far away form the smallest scales which are related to the
dissipation processes. There are at least three reasons for this
interest: First of all, it is well-known from experimental and also
theoretical studies that the deviations from the statements of the
famous classical Kolmogorov-Obukhov phenomenological theory (see,
e.g., \cite{MonYagBook,McComb,Frisch}) is surprisingly more
noticeable and visible for the simpler models of passively advected
scalar quantity (scalar field) than for the velocity field itself
(see, e.g.,
\cite{AnHoGaAn84,Sre91,HolSig94,Pumir94,TonWar94,ElKlRo90te});
second, at the same time, the problem of passive advection of scalar
field (as well as vector field) is much easier from theoretical
point of view than the original problem of anomalous scaling in the
framework of Navier-Stokes velocity field, and, in the end, third,
even very simplified models with given Gaussian statistics of
velocity field lead to the anomalous behavior which describe many
features of the real turbulent advection, see, e.g.,
Refs.\,\cite{Kra68,HolSig94,Pumir94,TonWar94,ElKlRo90te,all,Shraiman,Pumir,AveMaj90,ZhaGli92,Kraichnan94,
KrYaCh95} and references cited therein. As was already mentioned,
the crucial role in these studies was played (and is still played)
by the aforementioned rapid change model of a passive scalar
advection, where, for the first time, the systematic analysis of the
corresponding anomalous exponents was done on the microscopic level.
For example, within the so-called "zero-mode approach" to the rapid
change model \cite{all} (see also survey paper \cite{FaGaVe01}) the
anomalous exponents are found from the homogenous solutions (zero
modes) of the closed equations for the single-time correlations.

On the other hand, one of the most effective approach for studying
self-similar scaling behavior is the method of the field theoretic
renormalization group (RG) \cite{ZinnJustin,Vasiliev}. It can be
also used in the theory of fully developed turbulence
\cite{deDoMa79,AdVaPi83} and related problems, e.g., in the problem
of a passive scalar advection by the turbulent environment
\cite{AdVaHn84} (see also Refs.\,\cite{Vasiliev,AdAnVa96,AdAnVa99}
for details).

In Refs.\,\cite{AdAnVa98+} the field theoretic RG and
operator-product expansion (OPE) were used in the systematic
investigation of the rapid-change model of a passive scalar. It was
shown that within the field theoretic approach the anomalous scaling
is related to the very existence of so-called "dangerous" composite
operators with negative critical dimensions in OPE (see, e.g.,
Refs.\,\cite{Vasiliev,AdAnVa99} for details). In the subsequent
papers \cite{AdAnBaKaVa01} the anomalous exponents of the model were
calculated within the $\varepsilon$ expansion up to order
$\varepsilon^3$ (three-loop approximation). Here $\varepsilon$ is a
parameter which describes a given equal-time pair correlation
function of the velocity field (see next section). Important
advantages of the RG approach are its universality and calculational
efficiency: a regular systematic perturbation expansion for the
anomalous exponents was constructed, similar to the well-known
$\epsilon$-expansion in the theory of phase transitions.

Besides various generalizations of the Kraichnan model towards more
realistic ones, namely, models with inclusion of small scale
anisotropy \cite{AdAnHnNo00}, compressibility \cite{AdAn98,AnHo01},
and finite correlation time of the velocity field
\cite{Antonov99,Antonov00,all100} were studied by the field
theoretic approach. General conclusion of all these investigations
is that the anomalous scaling, which is the most intriguing and
important feature of the Kraichnan rapid change model, remains valid
for all generalized models.

In Ref.\,\cite{AdAnHnNo00} the field theoretic RG and OPE were
applied to the rapid change model of passive scalar advected by
Gaussian strongly anisotropic velocity field where the anomalous
exponents of the structure functions were calculated to the first
order in $\varepsilon$ expansion. It was shown that in the presence
of small-scale anisotropy the corresponding exponents are
nonuniversal, i.e., they are functions of the anisotropy parameters,
and they form the hierarchy with the leading exponent related to the
most "isotropic" operator. The importance of these investigations is
related to the question of the influence of anisotropy on
inertial-range behavior of passively advected fields as well as the
velocity field itself
\cite{Shraiman,Pumir,Antonov99,Antonov00,all101,all102,alll,SaVe94BoOr96,all5,YoKa01}
(see also the survey paper \cite{BiPr05} and references cited
therein, as well as recent astrophysical investigations, e.g, in
Refs.\,\cite{BiBiGaVe06,SoCaBrVe06}). On one hand, it was shown that
for the even structure (or correlation) functions the exponents
which describe the inertial-range scaling exhibit universality and
they are ordered hierarchically in respect to degree of anisotropy
with leading contribution given by the exponent from the isotropic
shell but, on the other hand, the survival of the anisotropy in the
inertial-range is demonstrated by the behavior of the odd structure
functions, namely, the so-called skewness factor decreases down the
scales slower than expected earlier in accordance with the classical
Kolmogorov-Obukhov theory.

On the other hand, in Ref.\,\cite{AnHo01} the influence of
compressibility and large-scale anisotropy on the anomalous scaling
behavior was studied in the aforementioned model in the order
$\varepsilon^2$ in the $\varepsilon$ expansion, and anomalous
exponents of higher-order correlation functions were calculated as
functions of the parameter of compressibility. It was shown that the
exponents exhibit hierarchy related to the degree of anisotropy.
Again, the existence of small-scale anisotropy effects were
demonstrated by the odd dimensionless ratios of correlation
functions: the skewness and hyperskewness factors, and it was shown
that the persistent of small-scale anisotropy is more pronounced for
larger values of the compressibility parameter $\alpha$.  From this
point of view, compressible systems are very interesting to be
studied.

In present paper, we shall continue in the investigation of the
model, namely, for the first time, we shall study the model with
compressible velocity field together with assumption about its
small-scale uniaxial anisotropy in the first order in $\varepsilon$,
i.e., we shall extend the model studied in Ref.\,\cite{AdAnHnNo00}
to the compressible case. In this situation, in general, two types
of diffusion-advection problems of scalar particles exist in the
compressible case which are identical in the incompressible case. In
what follows, we shall study only one of them, namely, the advection
of so-called "tracer" (see, e.g., Ref.\, \cite{AnHo01}), i.e., for
example, concentration of a scalar impurity, temperature, entropy,
etc. Details will be shown in the next section.

First of all we shall establish stability of the scaling regime of
the model and coordinates of the corresponding infrared (IR) stable
fixed point will be found analytically as functions of the
compressibility and anisotropy parameters. These results will be
then used in the analysis of the asymptotic behavior of the
single-time structure functions of a passively advected scalar
field.

The paper is organized as follows: In Sec.\,\ref{sec:Model} we
present definition of the model and introduce the compressibility
and small-scale uniaxial anisotropy to the given pair correlation
function of the velocity field. In Sec.\,\ref{sec:Field} we give the
field theoretic formulation of the original stochastic problem and
discuss the corresponding diagrammatic technique. The analysis of
the ultraviolet (UV) divergences of the model is given, the
multiplicative renormalizability of the model is established, and
the renormalization constants are calculated in one-loop
approximation. In Sec.\,\ref{sec:ScalReg} we analyze infrared (IR)
asymptotic behavior of the model which is governed by the IR stable
fixed point. Explicit expressions for the coordinates of the IR
fixed point are found.  In Sec.\,\ref{sec6} the renormalization of
needed composite operators is done and their critical dimensions are
found as functions of parameters of the model. In
Sec.\,\ref{sec:Conc} discussion of results is present.

\section{The model of advection of passive "tracer"} \label{sec:Model}

We shall consider the problem of the advection of a passive "tracer"
$\theta(x)\equiv \theta(t,{\bf x})$ which is described by the
following stochastic equation
\begin{equation}
\partial_t \theta = \nu_0 \triangle \theta - (v_i
\partial_i) \theta + f, \label{stoch1}
\end{equation}
where $\partial_t \equiv \partial/\partial t$, $\partial_i\equiv
\partial/\partial x_i$, $\triangle\equiv \partial^2$ is
the Laplace operator, $\nu_0$ is the coefficient of molecular
diffusivity (hereafter all parameters with a subscript $0$ denote
bare parameters of unrenormalized theory; see below), $v_i \equiv
v_i(x)$ is the $i$-th component of the compressible velocity field
${\bf v}(x)$, and $f \equiv f(x)$ is a Gaussian random noise with
zero mean and correlation function
\begin{equation}
D^f \equiv\langle f(x) f(x^{\prime})\rangle =
\delta(t-t^{\prime})C({\bf r}/L), \,\,\, {\bf r}={\bf x}-{\bf
x^{\prime}},\label{correlator}
\end{equation}
where parentheses $\langle...\rangle$ hereafter denote average over
corresponding statistical ensemble. The noise defined in
Eq.\,(\ref{correlator}) maintains the steady-state of the system but
the concrete form of the correlator will not be essential in what
follows. The only condition which must be fulfilled by the function
$C({\bf r}/L)$ is that it must decrease rapidly for $r\equiv |{\bf
r}| \gg L$, where $L$ denotes an integral scale related to the
stirring.

As was already mentioned in Introduction, another type of the
diffusion-advection problem exists in the case when compressibility
of the velocity field is supposed which is given by the following
more general stochastic equation \cite{AnHo01,LanLif87}
\begin{equation}
\partial_t \theta = \nu_0 \triangle \theta - \partial_i(v_i \theta) + f. \label{stoch2}
\end{equation}
It describes the passive advection of a density field but, in what
follows, we shall not study this problem. In the case of
incompressible velocity field (it is given mathematically by the
divergence-free condition $\partial_i v_i=0$) both models are
equivalent.

In real problems it is traditionally assumed that the velocity field
${\bf v}(x)$ satisfies stochastic Navier-Stokes equation
\cite{AdVaHn84}. In spite of, in what follows, we shall suppose that
the velocity field obeys a Gaussian distribution with zero mean and
two-point correlator
\begin{equation}
\hspace{-1cm} \langle v_i(x) v_j(x^{\prime}) \rangle \equiv
D^v_{ij}(x; x^{\prime})=  D_0 \delta(t-t^{\prime}) \int \frac{d^d
k}{(2\pi)^{d}} \frac{R_{ij}({\bf k})}{(k^2+m^2)^{d/2+\varepsilon}}
\exp[i{\bf k}({\bf x}-{\bf x^{\prime}})], \label{corv}
\end{equation}
i.e, we shall work in the framework of the so-called rapid-change
model
\cite{Kra68,all,Kraichnan94,KrYaCh95,AdAnVa98+,AdAnBaKaVa01,AdAnHnNo00,AdAn98,Obu49,Eyink96,Kraichnan97}.
Here, $d$ denotes the dimension of the ${\bf x}$ space, and $D_0$ is
an amplitude factor related to the coupling constant $g_0$ of the
model (expansion parameter in the perturbation theory, see next
Section) by the relation $D_0/\nu_0\equiv
g_0\simeq\Lambda^{2\varepsilon}$, where $\Lambda$ is the
characteristic UV momentum scale.  The parameter of the energy
spectrum of the velocity field $0<\varepsilon<1$ is taken in such a
way that its "Kolmogorov" value (the value which corresponds to the
Kolmogorov scaling of the velocity correlation function in developed
turbulence) is $\varepsilon=2/3$, and $1/m$ is another integral
scale. In general, the scale $1/m$ may be different from the
integral scale $L$ introduced in Eq.\,(\ref{correlator}) but,  in
accordance with Ref.\,\cite{AdAnHnNo00}, we suppose that $1/m\simeq
L$.

In the incompressible isotropic case the second-rank tensor
$R_{ij}({\bf k})$ in Eq.\,(\ref{corv}) has the simple form of the
ordinary transverse projector $R_{ij}({\bf k})=P_{ij}({\bf k})\equiv
\delta_{ij}-k_i k_j/k^2$, where $k=|{\bf k}|$. This tensor is
changed when one incorporates small-scale anisotropy or
compressibility. Let us briefly discuss these questions.

First of all, in the case of incompressible anisotropic case a new
second-rank tensor must be again a transverse operator because the
velocity field is still divergence-free ($\partial_i v_i=0$). The
simplest way how to introduce small-scale uniaxial anisotropy is to
take the operator $R_{ij}({\bf k})$ in the following way
\cite{AdAnHnNo00}
\begin{equation}
R_{ij} ({\bf k})  =
\left(1 + \alpha_{1} \frac {({\bf n \cdot k})^2} {k^2}\right) P_{ij}
({\bf k}) + \alpha_{2} n_s n_l P_{is} ({\bf k}) P_{jl} ({\bf k})\,,
\label{T-ij}
\end{equation}
where $P_{ij} ({\bf k})$ is the usual transverse projection operator
(as defined above), the unit vector ${\bf n}$ determines the
distinguished direction, and $\alpha_{1}$, $\alpha_{2}$ are
parameters characterizing the anisotropy. The positive definiteness
of the correlation function (\ref{corv}) imposes the following
restrictions on their values: $\alpha_{1}\,,\alpha_{2}>-1$. The
operator (\ref{T-ij}) is a special case of the general transverse
structure that possesses uniaxial anisotropy:
\begin{equation}
R_{ij}({\bf k})= a(\psi) P_{ij}({\bf k}) + b(\psi) n_s n_l P_{is}
({\bf k}) P_{jl}({\bf k})\,, \label{generalT}
\end{equation}
where $\psi$ denotes the angle between the vectors ${\bf n}$ and
${\bf k}$ (${\bf n \cdot k}=k \cos \psi$). Using Gegenbauer
polynomials \cite{GradshtejnRyzhik} the scalar functions in
representation (\ref{generalT}) may be expressed in the form
\[
a(\psi)=\sum_{i=0}^{\infty} a_{i} P_{2i}(\cos \psi)\,,\,\,\,\,
b(\psi)=\sum_{i=0}^{\infty} b_i P_{2i}(\cos \psi)\,.
\]
It was shown in Ref.\,\cite{AdAnHnNo00} that all main features of
the general model with the anisotropy structure represented by
Eq.\,(\ref{generalT}) are included in the simplified model with the
special form of the transverse operator given by Eq.\,(\ref{T-ij}).

Second, in the case of the compressible isotropic velocity field the
second-rank tensor $R_{ij}({\bf k})$ in Eq.\,(\ref{corv}) is not
longer a transverse projector  but it is rather a combination of the
transverse projector $P_{ij}({\bf k})\equiv\delta_{ij}-k_i k_j/k^2$
and of the longitudinal projector $Q_{ij}({\bf k})\equiv k_i
k_j/k^2$ as a result of the fact that the velocity field is not
longer solenoidal one. Thus, in this case, the correlator
(\ref{corv}) contains the tensor structure of the following general
form
\begin{equation}
R_{ij}({\bf k})=P_{ij}({\bf k})+\alpha Q_{ij}({\bf k}) \label{comp}
\end{equation}
where $\alpha \geq 0$ is a free parameter of the compressibility.
The value $\alpha=0$ corresponds to the divergence-free
(incompressible) advecting velocity field. It was used in
Refs.\,\cite{AdAn98,AnHo01} for determination of the influence of
the compressibility on the anomalous scaling of the correlation
functions of scalar density, as well as tracer fields in two-loop
level.

In what follows, we shall study the combined effects given by the
small-scale anisotropy and compressibility and our aim will be to
study possible deviations from the conclusions given in
Ref.\,\cite{AdAnHnNo00}, where the rapid-change model of passively
advected scalar field with small scale uniaxial anisotropy was
studied. To do this, it is necessary to introduce into the velocity
field correlator (\ref{corv}) corresponding second-rank tensor which
will have needed properties. The simplest way how to do this is to
add longitudinal projector $Q_{ij}({\bf k})\equiv k_i k_j/k^2$ to
the uniaxial anisotropic transverse tensor structure given in
Eq.\,(\ref{T-ij}). Thus, the result tensor structure which will be
used in our investigations is defined as
\begin{equation}
R_{ij} ({\bf k}) = P_{ij} ({\bf k}) + \alpha Q_{ij} ({\bf k}) +
\alpha_{1} \frac {({\bf n \cdot k})^2} {k^2} P_{ij} ({\bf k}) +
\alpha_{2} n_s n_l P_{is} ({\bf k}) P_{jl} ({\bf k})\,.
\label{Tg-ij}
\end{equation}
It means that we have introduced the compressible term to the
isotropic component of the axially anisotropic tensor structure
(\ref{T-ij}) only. Therefore, the contributions of compressibility
and anisotropy are given by a simple sum of the corresponding terms.
Nevertheless, as we shall see, the combined effects of anisotropy
and compressibility on the results will not be a simple sum of them.

\section{\label{sec:Field}Field Theoretic Formulation of the Model, UV Renormalization and RG analysis}

According to the well-known general theorem (see, e.g.,
Refs.\cite{ZinnJustin,Vasiliev}) the stochastic problem
(\ref{stoch1}) and (\ref{correlator}) is equivalent to the
field-theoretic model of the set of three fields $\Phi \equiv
\{\theta, \theta^{\prime}, {\bf v}\}$ with the following action
functional
\begin{eqnarray}
S(\Phi)&=&-\frac{1}{2} \int dt_1\,d^d{\bf x_1}\,dt_2\,d^d{\bf x_2}
\, v_i(t_1,{\bf x_1}) [D_{ij}^v(t_1,{\bf
x_1};t_2,{\bf x_2})]^{-1} v_j(t_2,{\bf x_2}) \nonumber  \\
&&+ \frac{1}{2} \int dt_1\,d^d{\bf x_1}\,dt_2\,d^d{\bf x_2} \,
\theta^{\prime}(t_1,{\bf x_1}) D^f(t_1,{\bf
x_1};t_2,{\bf x_2}) \theta^{\prime}(t_2,{\bf x_2}) \nonumber \\
&&+ \int dt\,d^d{\bf x}\,\, \theta^{\prime}\left[-\partial_t -
v_i\partial_i+\nu_0\triangle \right]\theta, \label{action1}
\end{eqnarray}
where $\theta^{\prime}$ is an auxiliary field, and all summations
over the vector indices are implied. The second and the third
integral in Eq.\,(\ref{action1}) correspond to the
Martin-Siggia-Rose action \cite{Martin} for the stochastic problem
(\ref{stoch1}), (\ref{correlator}) at fixed velocity field ${\bf
v}$, and the first integral describes the Gaussian averaging over
${\bf v}$ defined by the correlator $D^v$ in Eq.\,(\ref{corv}) with
tensor $R_{ij} ({\bf k})$ given in Eq.\,(\ref{Tg-ij}).

\begin{figure}[t]
     \vspace{-0cm}
       \begin{flushleft}
       \leavevmode
       \epsfxsize=6.0cm
       \epsffile{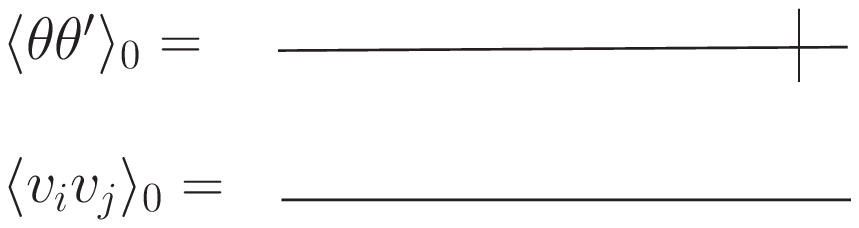}
   \end{flushleft}
     \vspace{-3.2cm}
   \begin{flushright}
      \leavevmode
       \epsfxsize=6.0cm
       \epsffile{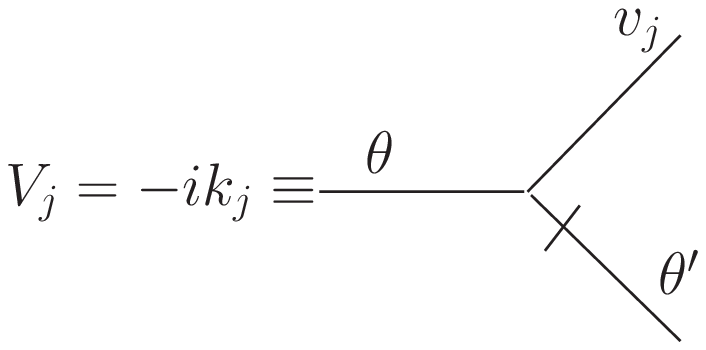}
   \end{flushright}
\vspace{-0.5cm} \caption{(Left) The graphical representation of
needed propagators of the model. (Right) The interaction vertex of
the model (wave-number-frequency representation). The momentum flows
into the vertex via the scalar field $\theta$. \label{fig1}}
\end{figure}

Action (\ref{action1}) is given in a form convenient for application
of the field theoretic perturbation analysis with the standard
Feynman diagrammatic technique. From the quadratic part of the
action one obtains the matrix of bare propagators. The
wave-number-frequency representation of propagators of the fields
$\theta$ and $\theta^{\prime}$ is
\begin{eqnarray}
\langle\theta \theta^{\prime}\rangle_0&=&\langle\theta^{\prime}
\theta\rangle^*_0=\frac{1}{-i\omega+\nu_0 k^2}, \label{thetathetap}\\
\langle\theta \theta \rangle_0&=& \frac{C({\bf k})}{(-i\omega+\nu_0
k^2)(i\omega+\nu_0 k^2)}, \label{thetatheta}\\
\langle\theta^{\prime} \theta^{\prime}\rangle_0 &=& 0
\end{eqnarray}
where $C({\bf k})$ is the Fourier transform of the function $C({\bf
r}/L)$ form Eq.\,(\ref{correlator}). On the other hand, the bare
propagator of the velocity field $\langle v v\rangle_0$ is defined
by Eq.\,(\ref{corv}) with the tensor structure given by
Eq.\,(\ref{Tg-ij}). In what follows, we shall need only the
propagators $\langle\theta \theta^{\prime}\rangle_0$ and $\langle v
v\rangle_0$. Their graphical representation is shown in
Fig.\,\ref{fig1}. The interaction in the model is given by the
nonlinear term $-\theta^{\prime} (v_i\partial_i)
\theta=\theta^{\prime} v_j V_j \theta$ with the vertex factor which
in the wave-number-frequency representation has the form (the
momentum flows into the vertex via the scalar field $\theta$):
\begin{equation}
V_j=-i k_j. \label{vertex}
\end{equation}
Its graphical representation is given in Fig.\,\ref{fig1}.

\input epsf

Standard power counting \cite{ZinnJustin,Vasiliev} leads to the
identification of correlation functions with superficial UV
divergences.  In the framework of the rapid-change passive advection
models detail analysis of this question was done, e.g., in
Ref.\,\cite{AdAnHnNo00}, where it was shown that the only
one-particle-irreducible (1PI) Green function which possesses
superficial UV-divergences is the function
$\Gamma_{\theta^{\prime}\theta}\equiv\langle \theta^{\prime} \theta
\rangle_{1-ir}$ (within the rapid-change model the situation is
unchanged when compressibility of the system is assumed) and, in the
isotropic case, this Green function leads only to the
renormalization of the term $\nu_0 \theta^{\prime} \triangle \theta$
of action (\ref{action1}) and the corresponding UV divergences may
be fully absorbed in the adequate redefinition of the existing
parameters $g_0$, $\nu_0$. Thus, all correlation functions
calculated in terms of the renormalized parameters $g,\nu$ are UV
finite.

The situation becomes, however, more complicated when anisotropy is
introduced. It is related to the fact that in this case the 1PI
Green function $\Gamma_{\theta^{\prime}\theta}$ contains divergences
corresponding to the structure $\theta^{\prime} ({\bf n}\cdot{\bf
\partial})^2\theta$ (the only possible anisotropic structure
$\sim \triangle\equiv \partial^2$) which is not present in original
unrenormalized action (\ref{action1}). It leads to the
non-renormalizability of the model in the anisotropic case. To make
the model multiplicatively renormalizable it is necessary to extend
original action (\ref{action1}) by including needed term with
corresponding new parameter. As a result the extended model is
described by the following action
\begin{eqnarray}
S(\Phi)&=&-\frac{1}{2} \int dt_1\,d^d{\bf x_1}\,dt_2\,d^d{\bf x_2}
\, v_i(t_1,{\bf x_1}) [D_{ij}^v(t_1,{\bf
x_1};t_2,{\bf x_2})]^{-1} v_j(t_2,{\bf x_2}) \nonumber  \\
&&+ \frac{1}{2} \int dt_1\,d^d{\bf x_1}\,dt_2\,d^d{\bf x_2} \,
\theta^{\prime}(t_1,{\bf x_1}) D^f(t_1,{\bf
x_1};t_2,{\bf x_2}) \theta^{\prime}(t_2,{\bf x_2}) \label{action2} \\
&&+ \int dt\,d^d{\bf x} \theta^{\prime}\left[-\partial_t -
v_i\partial_i+\nu_0\triangle  + \chi_0 \nu_0 ({\bf n}\cdot{\bf
\partial})^2\right]\theta,
\nonumber
\end{eqnarray}
where a new unrenormalized parameter $\chi_0$ has been introduced.
As was pointed out in Ref.\,\cite{AdAnHnNo00} the stability of the
system requires the positivity of the total viscous contribution
$\nu_0 k^2+\chi_0\nu_0 ({\bf n}\cdot{\bf k})^2$, i.e., the
inequality $\chi_0>-1$ must be fulfilled. This modification leads,
of course, also to the modification of the corresponding isotropic
propagators of the fields $\theta, \theta^{\prime}$ given in
Eqs.\,(\ref{thetathetap}) and (\ref{thetatheta}) which are now
defined as (see also Ref.\,\cite{AdAnHnNo00})
\begin{eqnarray}
\langle\theta \theta^{\prime}\rangle_0=\langle\theta^{\prime}
\theta\rangle^*_0&=&\frac{1}{-i\omega+\nu_0 k^2+\chi_0\nu_0 ({\bf n}\cdot{\bf k})^2}, \label{thetathetapa}\\
\langle\theta \theta \rangle_0&=& \frac{C({\bf k})}{|-i\omega+\nu_0
k^2+\chi_0\nu_0 ({\bf n}\cdot{\bf k})^2|^2}. \label{thetathetaa}
\end{eqnarray}

After this modifications the model defined by action (\ref{action2})
becomes multiplicatively renormalizable and the standard RG analysis
can be now applied. The corresponding renormalized action has the
form
\begin{eqnarray}
S^R(\Phi)&=&-\frac{1}{2} \int dt_1\,d^d{\bf x_1}\,dt_2\,d^d{\bf x_2}
\, v_i(t_1,{\bf x_1}) [D_{ij}^v(t_1,{\bf x_1};t_2,{\bf x_2})]^{-1} v_j(t_2,{\bf x_2}) \nonumber  \\
&&+ \frac{1}{2} \int dt_1\,d^d{\bf x_1}\,dt_2\,d^d{\bf x_2} \,
\theta^{\prime}(t_1,{\bf x_1}) D^f(t_1,{\bf
x_1};t_2,{\bf x_2}) \theta^{\prime}(t_2,{\bf x_2}) \label{action3} \\
 &&  +\int dt\,d^d{\bf x}
\theta^{\prime}\left[-\partial_t - v_i\partial_i+\nu Z_1 \triangle +
\chi \nu Z_2 ({\bf n}\cdot{\bf
\partial})^2\right]\theta,
\nonumber
\end{eqnarray}
where $Z_1$ and $Z_2$ are the renormalization constants (they absorb
the UV divergent parts of the 1PI function
$\Gamma_{\theta^{\prime}\theta}$). It is equivalent to the
multiplicative renormalization of the bare parameters $g_0,\nu_0$,
and $\chi_0$, namely
\begin{equation}
\nu_0=\nu Z_{\nu}, \quad g_0=g \mu^{2\varepsilon} Z_g,\quad
\chi_0=\chi Z_{\chi}, \label{zetka}
\end{equation}
where $g,\nu$, and $\chi$ are renormalized counterparts of the
corresponding bare parameters, and $\mu$ is a scale setting
parameter or the reference mass (it has the same canonical dimension
as the wave number). In what follows, we shall work in minimal
subtraction scheme (MS), therefore, in one-loop approximation, the
renormalization constants $Z$ have the form
$1+A(g,\alpha,\chi,\alpha_1,\alpha_2,d)/\varepsilon$. Thus, $A$ is a
function of dimensionless parameters but it is independent of
$\varepsilon$.

By comparison of the corresponding terms in action (\ref{action3})
with definitions of the renormalization constants $Z$ for parameters
(\ref{zetka}), one obtains the following relations between
renormalization constants:
\begin{equation}
Z_{\nu}=Z_1,\quad Z_{\chi}=Z_2 Z_1^{-1}, \quad
Z_g=Z_1^{-1},\label{zetka1}
\end{equation}
where last relation in Eq.\,(\ref{zetka1}) is a consequence of the
fact that $D_0$ defined  in Eq.\,(\ref{corv}) is not renormalized,
i.e., $D_0=g_0 \nu_0=g \nu \mu^{2\varepsilon}$.

Standardly, the formulation through the action functional
(\ref{action1}) (or through (\ref{action3}) in the anisotropic case)
replaces the statistical averages of random quantities in the
stochastic problem defined by Eqs.\,(\ref{stoch1}) and
(\ref{correlator}) with equivalent functional averages with weight
$\exp S(\Phi)$. Generating functionals of total Green functions G(A)
and connected Green functions W(A) are then defined by the
functional integral
\begin{equation}
G(A)=e^{W(A)}=\int {\cal D}\Phi \,\, e^{S(\Phi) +
A\Phi},\label{green}
\end{equation}
where $A(x)=\{A^{\theta},A^{\theta^{\prime}},{\bf A^{v}}\}$
represents a set of arbitrary sources for the set of fields $\Phi$,
${\cal D}\Phi \equiv {\cal D}\theta{\cal D}\theta^{\prime}{\cal
D}{\bf v}$ denotes the measure of functional integration, and linear
form $A\Phi$ is defined as
\begin{equation}
A\Phi= \int d\,x
[A^{\theta}(x)\theta(x)+A^{\theta^{\prime}}(x)\theta^{\prime}(x) +
A_i^{v}(x) v_i(x)].\label{form}
\end{equation}

Let us continue with renormalization of the model. The relation
$S(\theta,\theta^{\prime},{\bf v},
e_0)=S^R(\theta,\theta^{\prime},{\bf v}, e, \mu)$, where $e_0$
stands for the complete set of bare parameters and $e$ stands for
renormalized one, leads to the relation $W(A, e_0)=W^R(A, e, \mu)$
for the generating functional of connected Green functions. By
application of the operator $\tilde{\cal{D}}_{\mu}\equiv\mu
\partial_{\mu}$ at fixed $e_0$ on both sides of the latest equation
one obtains the basic RG differential equation
\begin{equation}
{\cal{D}}_{RG} W^R(A,e,\mu)=0, \label{RGE}
\end{equation}
where ${\cal{D}}_{RG}$ represents operation $\tilde{\cal{D}}_{\mu}$
written in the renormalized variables. Its explicit form is
\begin{equation}
{\cal{D}}_{RG} = {\cal{D}}_{\mu} +
\beta_g(g,\chi)\partial_g+\beta_{\chi}(g,\chi)\partial_u-\gamma_{\nu}(g,\chi){\cal{D}}_{\nu},\label{RGoper}
\end{equation}
where we denote ${\cal{D}}_x\equiv x\partial_x$ for any variable $x$
and the RG functions (the $\beta$ and $\gamma$ functions) are given
by well-known definitions (see, e.g.,
Refs.\,\cite{ZinnJustin,Vasiliev}) and, in our case, by using
relations (\ref{zetka1}) for renormalization constants, they have
the following form
\begin{eqnarray}
\gamma_{i}&\equiv& \tilde{\cal{D}}_{\mu} \ln Z_{i},\,\,\,i=1,2 \label{gammanu}\\
\beta_g&\equiv&\tilde{\cal{D}}_{\mu} g =g
(-2\varepsilon+\gamma_{1}), \label{betag}\\
\beta_{\chi}&\equiv&\tilde{\cal{D}}_{\mu} \chi =\chi
(\gamma_1-\gamma_2).\label{betau}
\end{eqnarray}

\begin{figure}[t]
     \vspace{0cm}
       \begin{center}
       \leavevmode
       \epsfxsize=6cm
       \epsffile{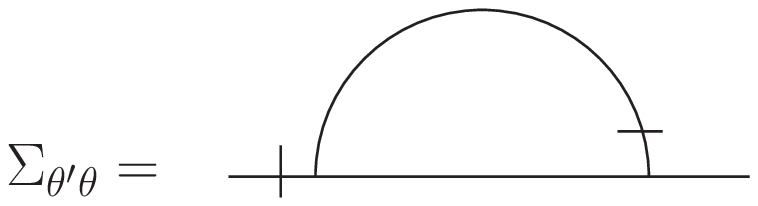}
   \end{center}
\vspace{-0.2cm} \caption{The only diagram which contribute to the
self-energy operator $\Sigma_{\theta^{\prime}\theta}$. \label{fig3}}
\end{figure}

The renormalization constants $Z_{1,2}$ are determined by the
requirement that the one-particle irreducible Green function
$\langle \theta^{\prime} \theta\rangle_{1-ir}$ must be UV finite
when is written in renormalized variables. In our case it means that
they have no singularities in the limit $\varepsilon\rightarrow0$.
The one-particle irreducible Green function $\langle \theta^{\prime}
\theta\rangle_{1-ir}$ is related to the self-energy operator
$\Sigma_{\theta^{\prime}\theta}$ by the Dyson equation
\begin{equation}
\langle \theta^{\prime}\theta \rangle_{1-ir}=-i\omega+\nu_0 p^2
+\nu_0\chi_0({\bf n}\cdot{\bf p})^2 -
\Sigma_{\theta^{\prime}\theta}(\omega, p).\label{Dyson}
\end{equation}
Thus $Z_{1,2}$ are found from the requirement that the UV
divergences are canceled in Eq.\,(\ref{Dyson}) after substitutions
$\nu_0=\nu Z_{1}$ and $\nu_0 \chi_0=\nu \chi Z_2$. This determines
$Z_{1,2}$ up to an UV finite contributions, which are fixed by the
choice of the renormalization scheme. In the MS scheme, as was
mentioned already above, all the renormalization constants have the
form: 1 + {\it pole in $\varepsilon$}. In one-loop approximation the
self-energy operator $\Sigma_{\theta^{\prime}\theta}$ is represented
by the corresponding one-particle irreducible diagram which is shown
in Fig.\,\ref{fig3} but it must be stressed that, at the same time,
it is an exact result because in rapid-change model all higher-loop
diagrams contain at least one closed loop which is built on by
retarded or advanced propagators only, thus they are automatically
equal to zero.

\input epsf

The diagram in Fig.\,\ref{fig3} has the following analytical
representation
\begin{equation}
\Sigma_{\theta^{\prime}\theta}=-\frac{S_d}{(2\pi)^d}\frac{g
\nu}{4}\left(\frac{\mu}{m}\right)^{2\varepsilon}\frac{1}{\varepsilon}(p^2
A_1+({\bf p}\cdot {\bf n})^2 A_2)\,, \label{sigmavys}
\end{equation}
where $S_d=2\pi^{d/2}/\Gamma(d/2)$ denotes the d-dimensional sphere,
and the functions $A_1$ and $A_2$ have the following simple explicit
form
\begin{eqnarray}
A_1&=&\frac{(d+2)(d-1+\alpha)+\alpha_1
(d+1)+\alpha_2}{d(d+2)}, \label{AA1}\\
A_2&=&\frac{-2(\alpha_1+\alpha_2)+d^2\alpha_2}{d(d+2)},\label{AA2}
\end{eqnarray}

In the end, the renormalization constants $Z_1$ and $Z_2$ are given
as follows
\begin{eqnarray}
Z_1=1-\frac{g}{\varepsilon}\frac{S_d}{(2\pi)^d}\frac{A_1}{4}\,, \label{zz1} \\
Z_2=1-\frac{g}{\varepsilon}\frac{S_d}{(2\pi)^d}\frac{A_2}{4\chi}\,.
\label{zz2}
\end{eqnarray}
Because of the second rank tensor (\ref{Tg-ij}) is taken in the
simple form of the sum of the uniaxial anisotropic transverse
projector and the longitudinal projector then, as a result of this
fact, the self-energy operator $\Sigma_{\theta^{\prime}\theta}$ is
also a simple sum of the results obtained in
Refs.\,\cite{AdAnHnNo00} and \cite{AnHo01}. Using the definition of
anomalous dimensions $\gamma_{1,2}$ in Eq.\,(\ref{gammanu}) one
comes to the following expressions
\begin{equation}
\gamma_1=\frac{\bar{g}}{2} A_1\,, \quad \label{gammas}
\gamma_2=\frac{\bar{g}}{2\chi} A_2,
\end{equation}
where we denote $\bar{g}=g S_d/(2\pi)^d$, and $A_{1,2}$ are given in
Eqs.\,(\ref{AA1}) and (\ref{AA2}).

Let us note once more that the expressions
(\ref{zz1})-(\ref{gammas}) are exact as a result of non-existence of
the higher-loop corrections.

\section{Fixed point and scaling regime}\label{sec:ScalReg}

Possible scaling regimes of a renormalizable model are directly
given by the IR stable fixed points of the corresponding system of
RG equations \cite{ZinnJustin,Vasiliev}. The fixed point of the RG
equations is defined by $\beta$-functions, namely, by requirement of
their vanishing. In our model, the coordinates $g_*, \chi_*$ of a
fixed point are found from the system of two equations
\begin{equation}
\beta_g(g_*,\chi_*)=\beta_{\chi}(g_*,\chi_*)=0.
\end{equation}
The beta functions $\beta_g$ and $\beta_{\chi}$ are defined in
Eqs.\,(\ref{betag}), and (\ref{betau}). To investigate the IR
stability of a fixed point it is enough to analyze the eigenvalues
of the matrix $\Omega$ of the first derivatives:
\begin{equation}
\Omega_{ij}=\left(\begin{array}{cc}\partial \beta_g/\partial g &
\partial \beta_g/\partial \chi \\ \partial \beta_{\chi}/\partial g & \partial \beta_{\chi}/\partial
\chi
\end{array}
\right).
\end{equation}
The IR asymptotic behavior is governed by the IR stable fixed
points, i.e., those for which both eigenvalues are positive. Using
the explicit expressions given in Eq.\,(\ref{gammas}) together with
the definitions of $\beta$ functions in Eqs.\,(\ref{betag}) and
(\ref{betau}) leads to the explicit expressions for the coordinates
of the non-trivial IR stable fixed point
\begin{eqnarray}
\bar{g}_*&=&\frac{4 d (d+2)\varepsilon}{(d+2)(d-1+\alpha)+\alpha_1
(d+1)+\alpha_2},\label{gAA1}
\\ \chi_*&=&
\frac{-2(\alpha_1+\alpha_2)+d^2\alpha_2}{(d+2)(d-1+\alpha)+\alpha_1
(d+1)+\alpha_2}.\label{hAA2}
\end{eqnarray}
In the incompressible limit $\alpha\rightarrow 0$ one comes to the
results of Ref.\,\cite{AdAnHnNo00}, namely
\begin{eqnarray}
\bar{g}_*^{a}&=&\frac{4 d (d+2)\varepsilon}{(d+2)(d-1)+\alpha_1
(d+1)+\alpha_2},\label{gAA10}
\\ \chi_*^{a}&=&
\frac{-2(\alpha_1+\alpha_2)+d^2\alpha_2}{(d+2)(d-1)+\alpha_1
(d+1)+\alpha_2}.\label{hAA20}
\end{eqnarray}
On the other hand, in isotropic limit $\alpha_{1,2}\rightarrow 0$
one has (see, e.g., Ref.\,\cite{Antonov00})
\begin{equation}
\bar{g}_*^{c}=\frac{4 d (d+2)\varepsilon}{(d+2)(d-1+\alpha)}.
\end{equation}
Thus, as one can see, our definition of the second rank tensor
(\ref{Tg-ij}) leads to the simple relation among
$\bar{g}_*,\bar{g}_*^{a}$, and $\bar{g}_*^{c}$, namely
\begin{equation}
\frac{1}{\bar{g}_*}=\frac{1}{\bar{g}_*^{a}}+\frac{1}{\bar{g}_*^{c}}.
\end{equation}

The values of anomalous dimensions $\gamma_1$ and $\gamma_2$ are
found exactly at fixed point and are defined as follows
\begin{equation}
\gamma_1^*=\gamma_2^*=2\varepsilon,\label{gammyfix}
\end{equation}
where $\gamma_{1,2}^*\equiv\gamma_{1,2}(g_*,\chi_*)$.

The matrix of the first derivatives taken at the fixed point has
simple  diagonal form
\begin{equation}
\Omega_{ij}=\left(\begin{array}{cc} 2\varepsilon & 0 \\ 0 &
2\varepsilon
\end{array}
\right),
\end{equation}
i.e., the fixed point is IR stable if $\varepsilon>0$. This is also
the only condition to have $g_*>0$ (together, of course, with
earlier discussed physical assumptions, namely: $\alpha_{1,2}>-1$
and $\alpha \geq 0$, see Sec.\,\ref{sec:Model}). The physical
condition $\chi_*>-1$ is also fulfilled without further restrictions
on the parameter space.

The issue of interest are especially multiplicatively renormalizable
equal-time two-point quantities $G(r)$ (see, e.g.,
\cite{AdAnHnNo00}). The example of such quantity are the equal-time
structure functions
\begin{equation}
S_{N}(r)\equiv\langle[\theta(t,{\bf x})-\theta(t,{\bf
x'})]^{N}\rangle,\,\, r=|{\bf x}-{\bf x^{\prime}}| \label{struc}
\end{equation}
in the inertial range specified by the inequalities $l\sim 1/\Lambda
<<r<<L=1/m$ ($l$ is an internal length). Here parentheses $\langle
\cdots \rangle$ mean functional average over fields $\Phi=\{\theta,
\theta', {\bf v}\}$ with weight $\exp S^R(\Phi)$.

First let us describe briefly IR scaling behavior in general on the
example of an equal-time function $G(r)$ which is multiplicatively
renormalizable. The IR scaling behavior of the function $G(r)$ (for
$r/l\gg 1$ and any fixed $r/L$)
\begin{equation}
G(r)\simeq \nu_0^{d^{\omega}_G} l^{-d_G} (r/l)^{-\Delta_G} R(r/L)
\label{frscaling}
\end{equation}
is related to the existence of IR stable fixed points of the RG
equations (see above). In Eq.\,(\ref{frscaling}) $d^{\omega}_G$ and
$d_G$ are corresponding canonical dimensions of the function $G$:
$d^{\omega}_G$ is the frequency dimension, and $d_G$ is the total
canonical dimension. They are related by the relation $d_G=d_G^k+2
d^{\omega}_G$, where $d_G^k$ is corresponding momentum canonical
dimension. The existence of two independent canonical dimensions is
related to the fact that our model belongs among two-scale dynamical
models (details see, e.g., in Refs.\,\cite{Vasiliev,AdAnVa99}). In
Eq.\,(\ref{frscaling}) $R(r/L)$ is so-called scaling function which
cannot be determined by RG equation (see, e.g., \cite{Vasiliev}),
and $\Delta_G$ is the critical dimension defined as
\begin{equation}
\Delta_G=d_G^k+\Delta_{\omega} d_G^{\omega} + \gamma_G^*.
\end{equation}
Here $\gamma_G^*$ is the fixed point value of the anomalous
dimension $\gamma_G\equiv \mu \partial_{\mu} \ln Z_G$, where $Z_G$
is renormalization constant of multiplicatively renormalizable
quantity $G$, i.e., $G=Z_G G^R$ \cite{Antonov00}, and
$\Delta_{\omega}=2-\gamma_{\nu}^*=2-\gamma_{1}^*$ is the critical
dimension of frequency with $\gamma_{1}^*=2\varepsilon$ as it is
shown in Eq.\,(\ref{gammyfix}).

Now, let us apply the above discussion to the inertial-range
analysis of the equal-time structure functions as defined in
Eq.\,(\ref{struc}). It is well-known that, in the isotropic case,
the odd functions $S_{2n+1}$ vanish, while for $S_{2n}$ simple
dimensional considerations give
\begin{equation}
S_{2n}(r)=  \nu_0^{-n}\, r^{2n}\, R_{2n} (r/l, r/L),
\label{strucdim}
\end{equation}
where  $R_{2n}$ are some functions of dimensionless variables.
First, the multiplicative renormalizability of the model leads to
the existence of differential RG equations for these structure
functions and their asymptotic behavior for $ r/l>>1$ and any fixed
$r/L$ is given by IR stable fixed point of the RG equations and the
structure functions can be written in the following form
\begin{equation}
S_{2n}(r)= \nu_0^{-n}\, r^{2n}\, (r/l)^{-\gamma_{n}} R_{2n} (r/L),
\quad r/l>>1 \label{strucdim2}
\end{equation}
with unknown scaling functions $R_{2n} (r/L)$. In the theory of
critical phenomena \cite{ZinnJustin,Vasiliev} the quantity
$\gamma_n$ is known as ``anomalous dimension'' and
$\Delta[S_{2n}]\equiv-2n + \gamma_{n}$ is termed the ``critical
dimension'' (see above), where $-2n$ is the corresponding
``canonical dimension''. In our case, $\gamma_n=2 n \varepsilon$
\cite{AdAnVa98+}, i.e., representation (\ref{strucdim2}) implies the
existence of a scaling in the IR region ($r/l>>1$, $r/L$ fixed) with
definite critical dimensions of all ``IR relevant'' parameters,
$\Delta[S_{2n}] =-2n(1- \varepsilon)$, $\Delta_{r}=-1$,
$\Delta_{L^{-1}}=1$ and fixed ``irrelevant'' parameters $\nu_0$ and
$l$ regardless of the form of the functions $ R_{2n} (r/L)$.

The second stage of RG analysis is associated with the investigation
of small $r/L$ behavior of the functions $ R_{2n}(r/L)$ in
Eq.\,(\ref{strucdim2}) using the OPE. It shows that, in the limit
$r/L\to 0$, the functions $ R_{2n} (r/L)$ have the asymptotic form
\begin{equation}
R_{2n} (r/L) = \sum_{F} C_{F}(r/L)\, (r/L)^{\Delta_n}, \label{ope}
\end{equation}
where $C_{F}$ are some coefficients regular in $r/L$ and the
summation is implied over certain renormalized composite operators
$F$  with critical dimensions $\Delta_n$ \cite{Vasiliev}. In the
case under consideration, the leading operators $F$ have the form
$F_n= (\partial_i \theta \partial_i \theta)^n$. In Sec.\,\ref{sec6}
we shall consider them in detail where the complete calculation of
the critical dimensions  of the composite operators $F_n$ will be
present for arbitrary values of $n$, $d$, $\alpha$ and
$\alpha_{1,2}$.

\section{Operator product expansion, Critical
dimensions of composite operators, and Anomalous
scaling}\label{sec6}

\input epsf
   \begin{figure}[t]
     \vspace{0cm}
       \begin{center}
       \leavevmode
       \epsfxsize=3.6cm
       \epsffile{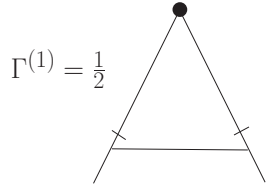}
   \end{center}
\vspace{0cm} \caption{Graphical representation of the one-loop
correction to the composite operator $\Gamma_N$ (details see in
Ref.\,\cite{AdAnHnNo00}). \label{fig4}}
\end{figure}

\subsection{Operator product expansion}

Let us now study the behavior of the scaling function in
Eq.\,(\ref{frscaling}). According to the OPE
\cite{ZinnJustin,Vasiliev,AdAnVa96,AdAnVa99}, the equal-time product
$F_1(x^{\prime})F_2(x^{\prime\prime})$ of two renormalized composite
operators \footnote{By definition we use the term "composite
operator" for any local monomial or polynomial constructed from
primary fields and their derivatives at a single point $x\equiv(t,
{\bf x})$. Constructions $\theta^n(x)$ and $[\partial_i \theta(x)
\partial_i \theta(x)]^n$ are typical examples} at ${\bf x}=({\bf x^{\prime}}+{\bf
x^{\prime\prime}})/2=const$ and ${\bf r}={\bf x^{\prime}}-{\bf
x^{\prime\prime}} \rightarrow 0$ can be written in the following
form
\begin{equation}
F_1(x^{\prime})F_2(x^{\prime\prime})=\sum_i C_{F_i}({\bf r})
F_i({\bf x},t), \label{fff}
\end{equation}
where summation is taken over all possible renormalized local
composite operators $F_i$ allowed by symmetry with definite critical
dimensions $\Delta_{F_i}$, and the functions $C_{F_i}$ are the
corresponding Wilson coefficients regular in $L^{-2}$. The
renormalized correlation function $\langle
F_1(x^{\prime})F_2(x^{\prime\prime}) \rangle$ can be now found by
averaging Eq.\,(\ref{fff}) with the weight $\exp S^R$ with $S^R$
from Eq.\,(\ref{action3}). The quantities $\langle F_i \rangle$
appear on the right-hand side, and their asymptotic behavior in the
limit $L^{-1} \rightarrow 0$ is then found from the corresponding RG
equations and has the form $\langle F_i \rangle \propto
L^{-\Delta_{F_i}}$.

From the OPE (\ref{fff}) one can find that the scaling function
$R(r/L)$ in the representation (\ref{frscaling}) for the correlation
function $F_1(x^{\prime})F_2(x^{\prime\prime})$ has the form
\begin{equation}
R(r/L)=\sum_{i} C_{F_i} (r/L)^{\Delta_{F_i}},\label{ff1}
\end{equation}
where the coefficients $C_{F_i}$ are regular in $(r/L)^2$.

The principal feature of the turbulence models is the existence of
the so-called "dangerous" operators with negative critical
dimensions \cite{Vasiliev,AdAnVa96,AdAnVa99,AdAnVa98+,AdAn98}. Their
presence in the OPE determines the IR behavior of the scaling
functions and leads to their singular dependence on $L$ when $r/L
\rightarrow0$. Therefore, the turbulence models are crucially
different from the models of critical phenomena, where the leading
contribution to the representation (\ref{ff1}) is given by the
simplest operator $F=1$ with the dimension $\Delta_F=0$, and the
other operators determine only the corrections that vanish for $r/L
\rightarrow0$.

\input epsf
   \begin{figure}[t]
     \vspace{-1cm}
       \begin{flushleft}
       \leavevmode
       \epsfxsize=8.5cm
       \epsffile{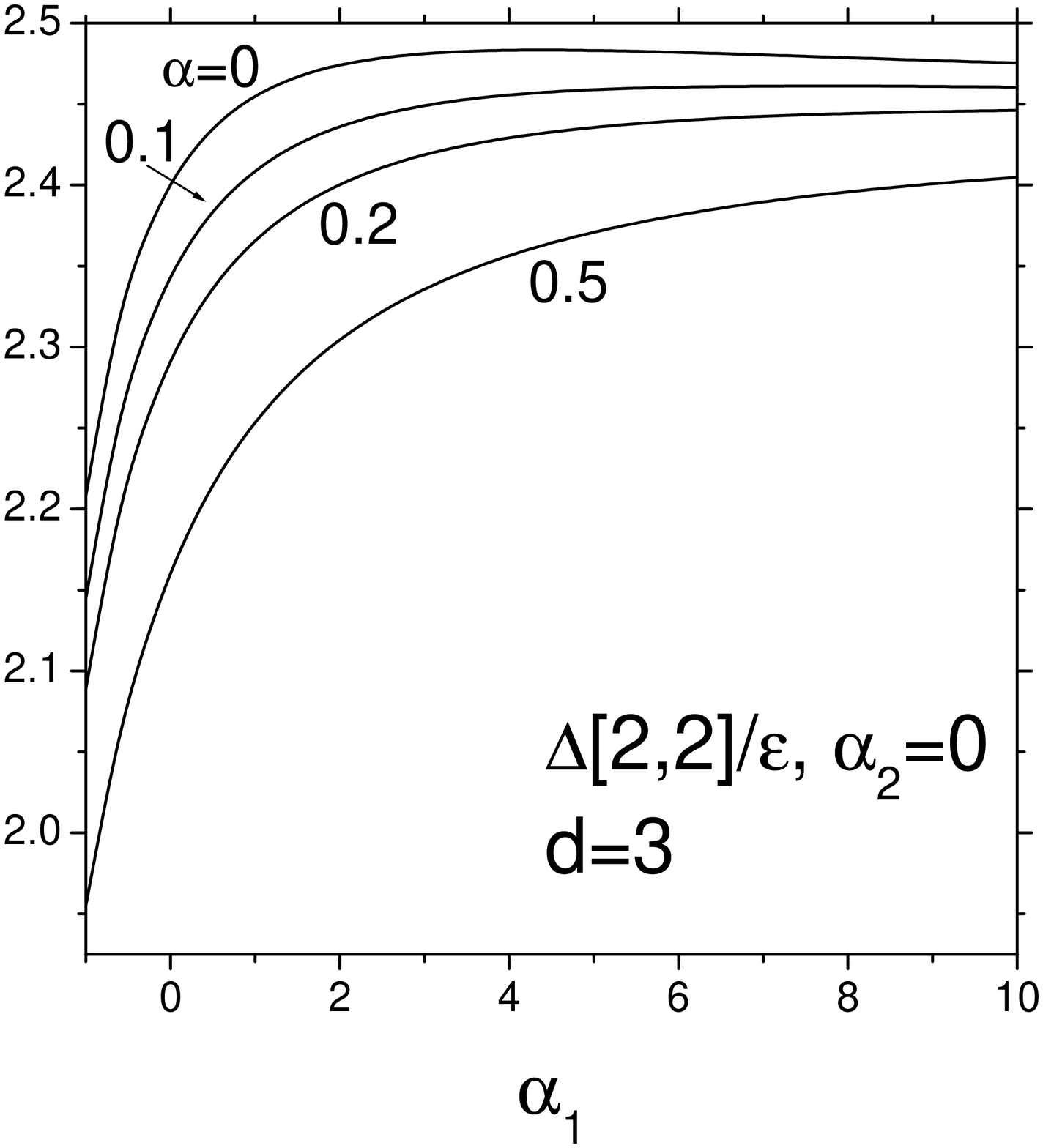}
   \end{flushleft}
     \vspace{-13.2cm}
   \begin{flushright}
       \leavevmode
       \epsfxsize=8.5cm
       \epsffile{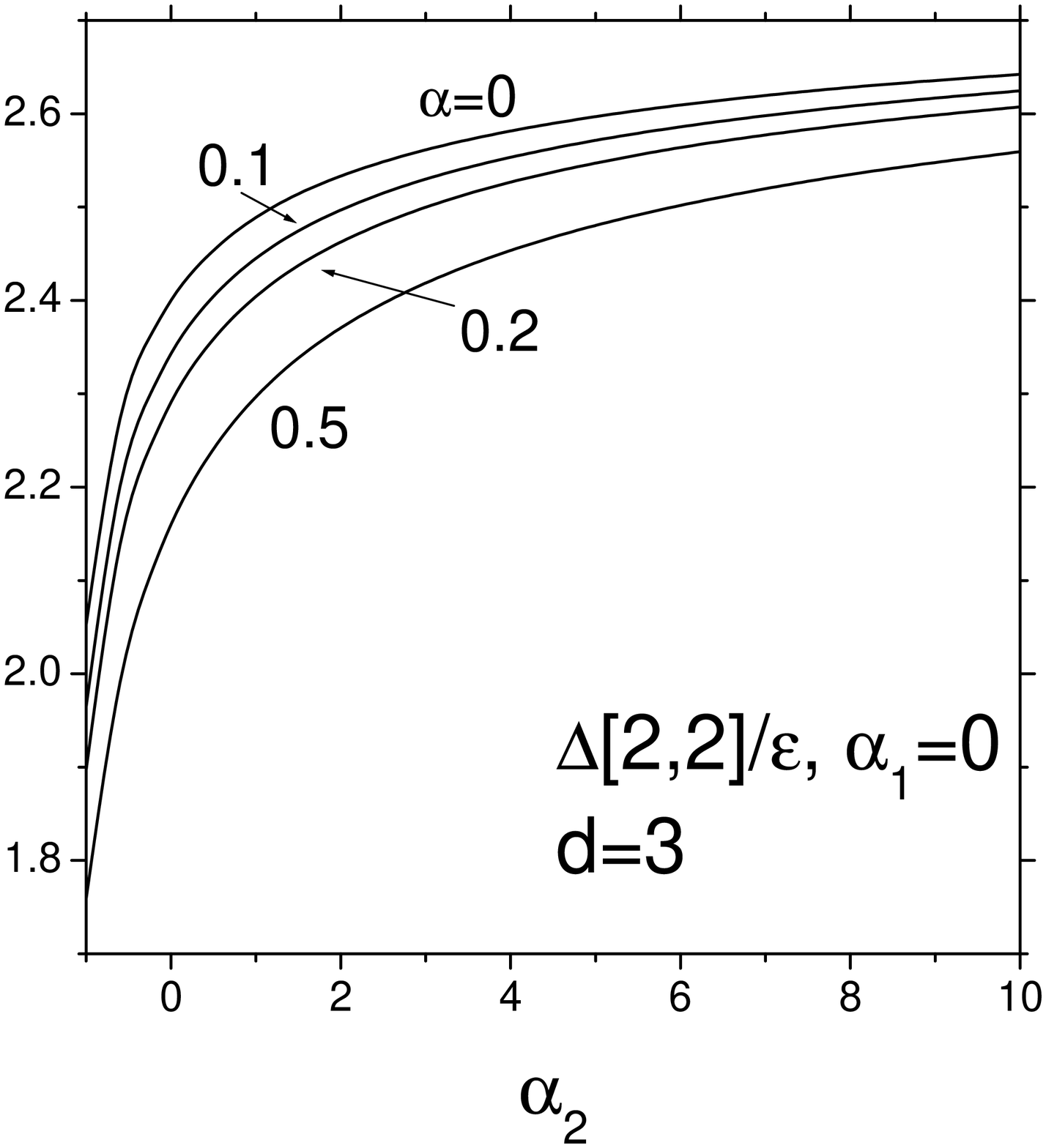}
   \end{flushright}
\vspace{-1.5cm} \caption{Dependence of the critical dimension
$\Delta[2,2]/\varepsilon$ on anisotropy parameters $\alpha_1$
($\alpha_2=0$) and $\alpha_2$ ($\alpha_1=0$) for different values of
the compressibility parameter $\alpha$. \label{fig56}}
\end{figure}

If the spectrum of the dimensions $\Delta_{F_i}$ for a given scaling
function is bounded from below, the leading term of its behavior for
$r/L \rightarrow 0$ is given by the minimal dimension. It will be
our case and for the investigation of the anomalous scaling of the
structure functions (\ref{struc}) the leading contribution of the
Taylor expansion is given by the tensor composite operators
constructed solely of the scalar gradients (see, e.g.,
Ref.\,\cite{AdAnHnNo00} for details)
\begin{equation}
F[N,p]\equiv \partial_{i_1}\theta \cdots \partial_{i_p}\theta
(\partial_{i}\theta
\partial_{i} \theta)^n, \label{composite}
\end{equation}
where $N=p+2n$ is the total number of the fields $\theta$ entering
into the operator and $p$ is the number of the free vector indices.

\input epsf
   \begin{figure}[t]
     \vspace{-1cm}
       \begin{flushleft}
       \leavevmode
       \epsfxsize=8.5cm
       \epsffile{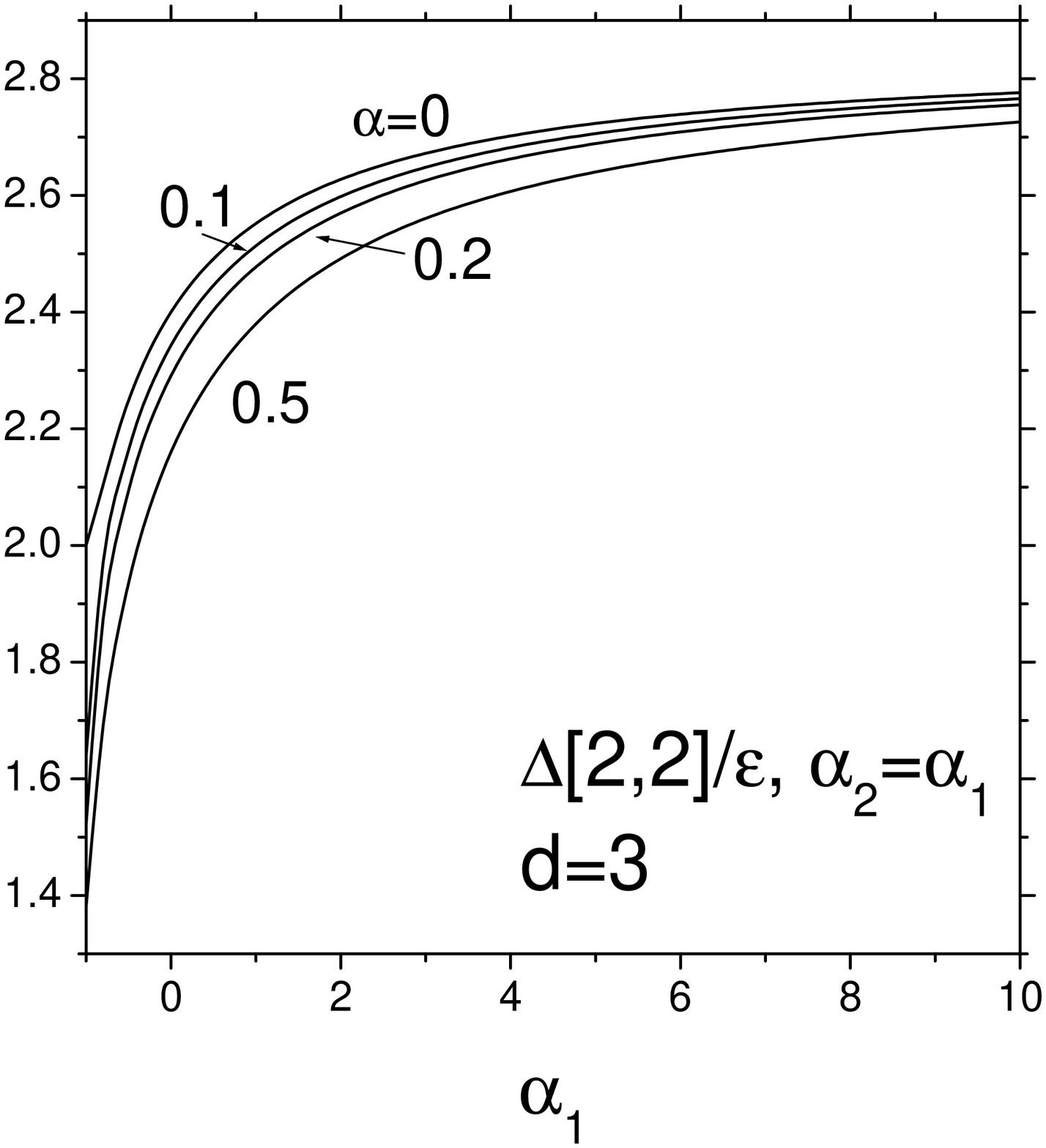}
   \end{flushleft}
     \vspace{-13.2cm}
   \begin{flushright}
       \leavevmode
       \epsfxsize=8.5cm
       \epsffile{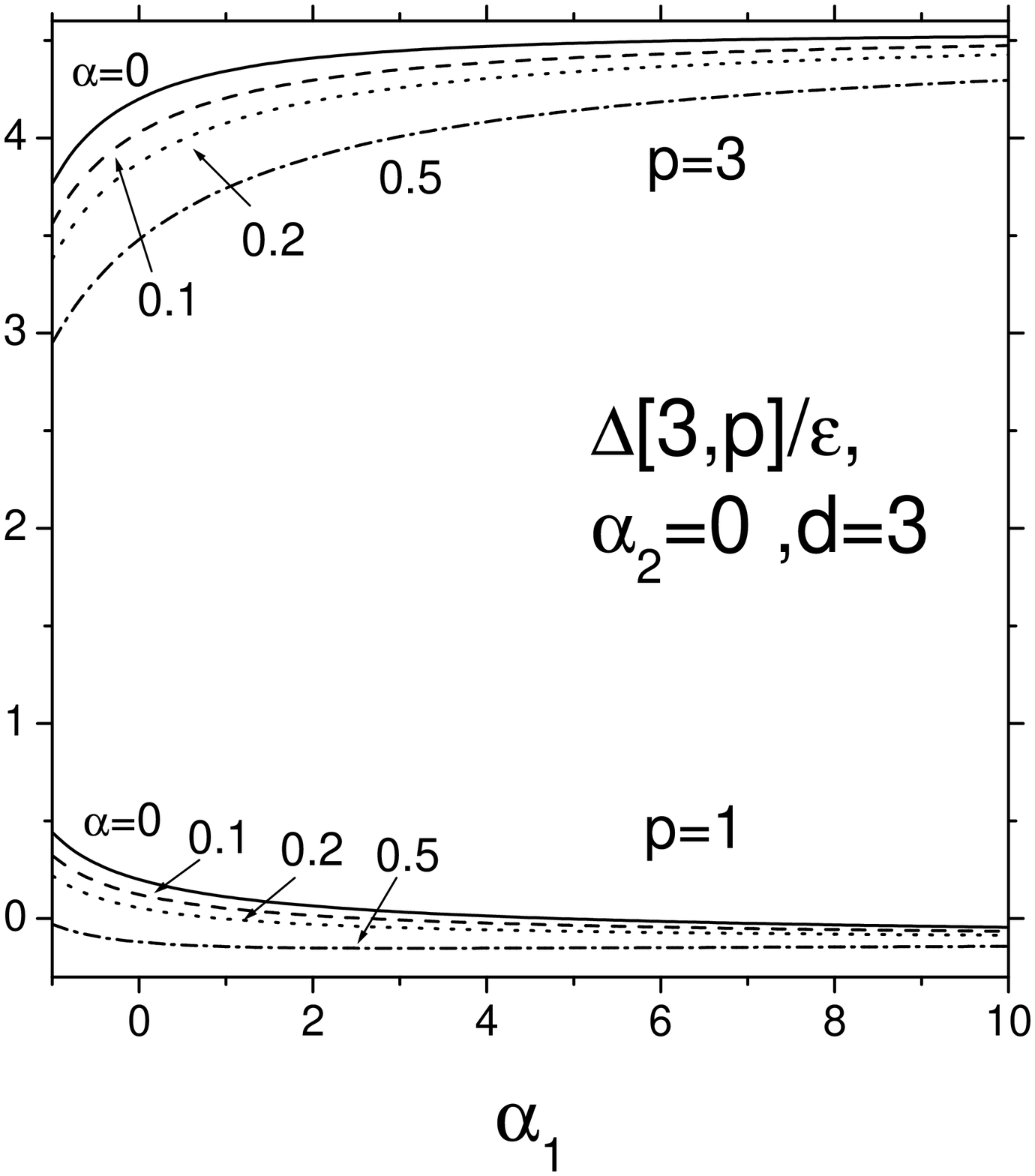}
   \end{flushright}
\vspace{-1.5cm} \caption{(Left) Dependence of the critical dimension
$\Delta[2,2]/\varepsilon$ on anisotropy parameters
$\alpha_1=\alpha_2$ for different values of the compressibility
parameter $\alpha$. (Right) Dependence of the critical dimension
$\Delta[3,p]/\varepsilon, p=1,3$ on anisotropy parameter $\alpha_1$
($\alpha_2=0$) for different values of the compressibility parameter
$\alpha$. \label{fig78}}
\end{figure}

\subsection{ Composite operators $F[N,p]$:
renormalization and critical dimensions}

Here we shall briefly discuss the renormalization of the composite
operators (\ref{composite}) which play central role in our
investigation (complete and detailed discussion of the
renormalization of this composite operators can be found in
Ref.\,\cite{AdAnHnNo00}) but first we describe the basis of general
theory.

\input epsf
   \begin{figure}[t]
     \vspace{-1cm}
       \begin{flushleft}
       \leavevmode
       \epsfxsize=8.5cm
       \epsffile{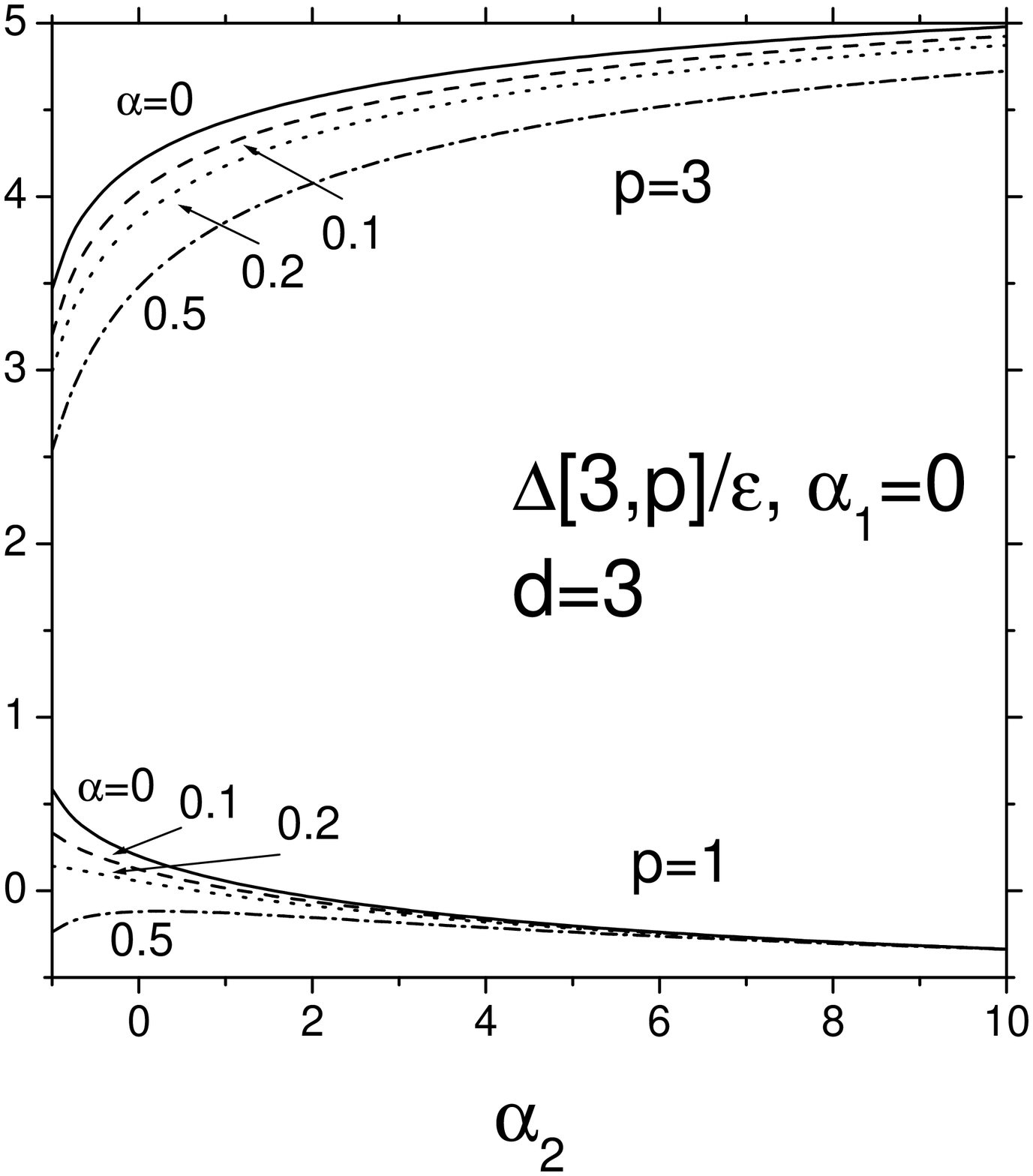}
   \end{flushleft}
     \vspace{-13.2cm}
   \begin{flushright}
       \leavevmode
       \epsfxsize=8.5cm
       \epsffile{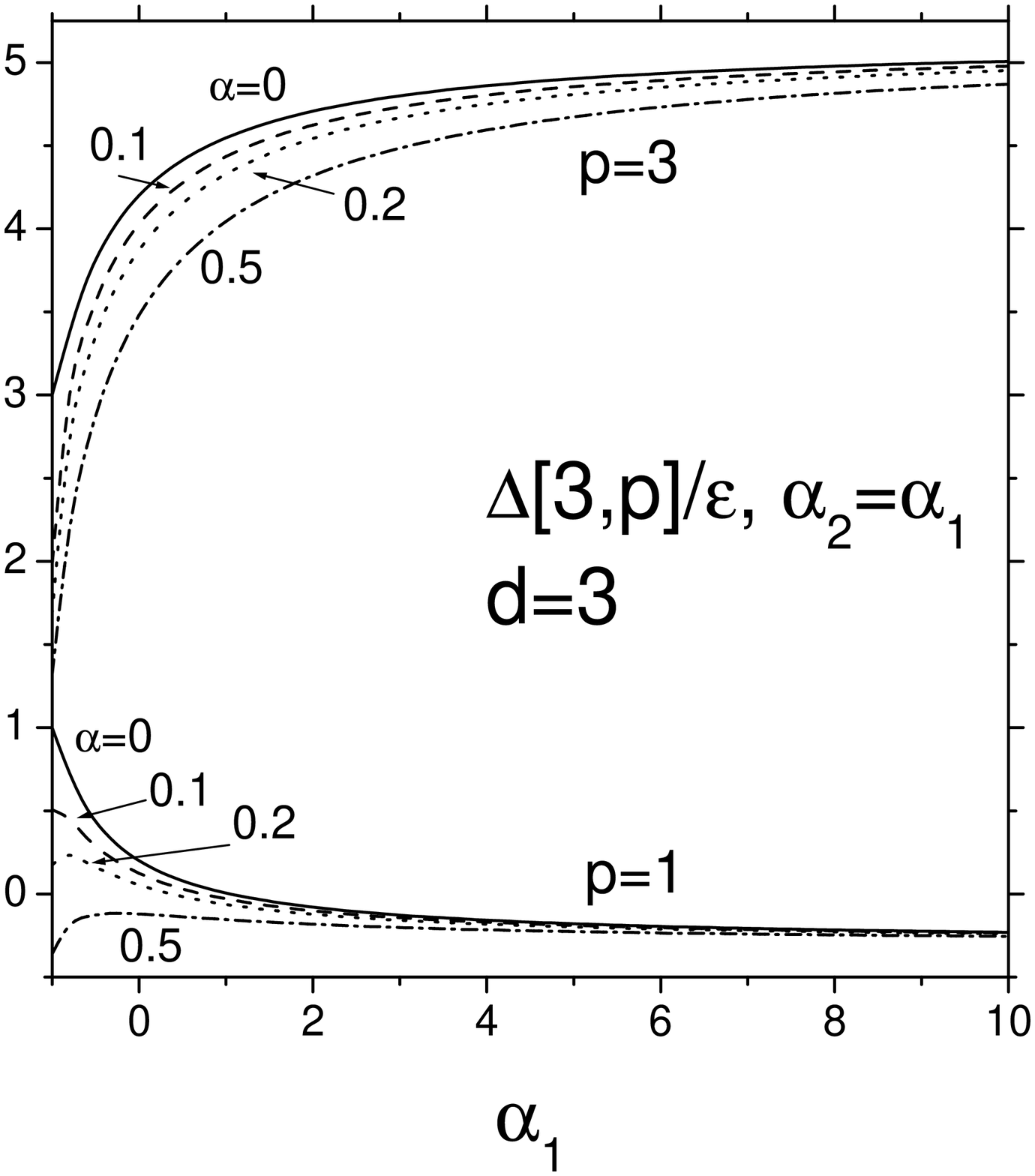}
   \end{flushright}
\vspace{-1.5cm} \caption{Dependence of the critical dimension
$\Delta[3,p]/\varepsilon, p=1,3$ on anisotropy parameters $\alpha_2$
($\alpha_1=0$) and $\alpha_1=\alpha_2$ for different values of the
compressibility parameter $\alpha$. \label{fig910}}
\end{figure}

The necessity of additional renormalization of the composite
operators (\ref{composite}) is related to the fact that the
coincidence of the field arguments in Green functions containing
them leads to additional UV divergences. These divergences must be
removed by special kind of renormalization procedure which can be
found, e.g., in Refs.\,\cite{ZinnJustin,Vasiliev,Collins}, where
their renormalization is studied in general. The renormalization of
composite operators in the models of turbulence is discussed in
Refs.\,\cite{AdVaPi83,AdAnVa99}. Besides, typically, the composite
operators are mixed under renormalization.

\input epsf
   \begin{figure}[t]
     \vspace{-1cm}
       \begin{flushleft}
       \leavevmode
       \epsfxsize=8.5cm
       \epsffile{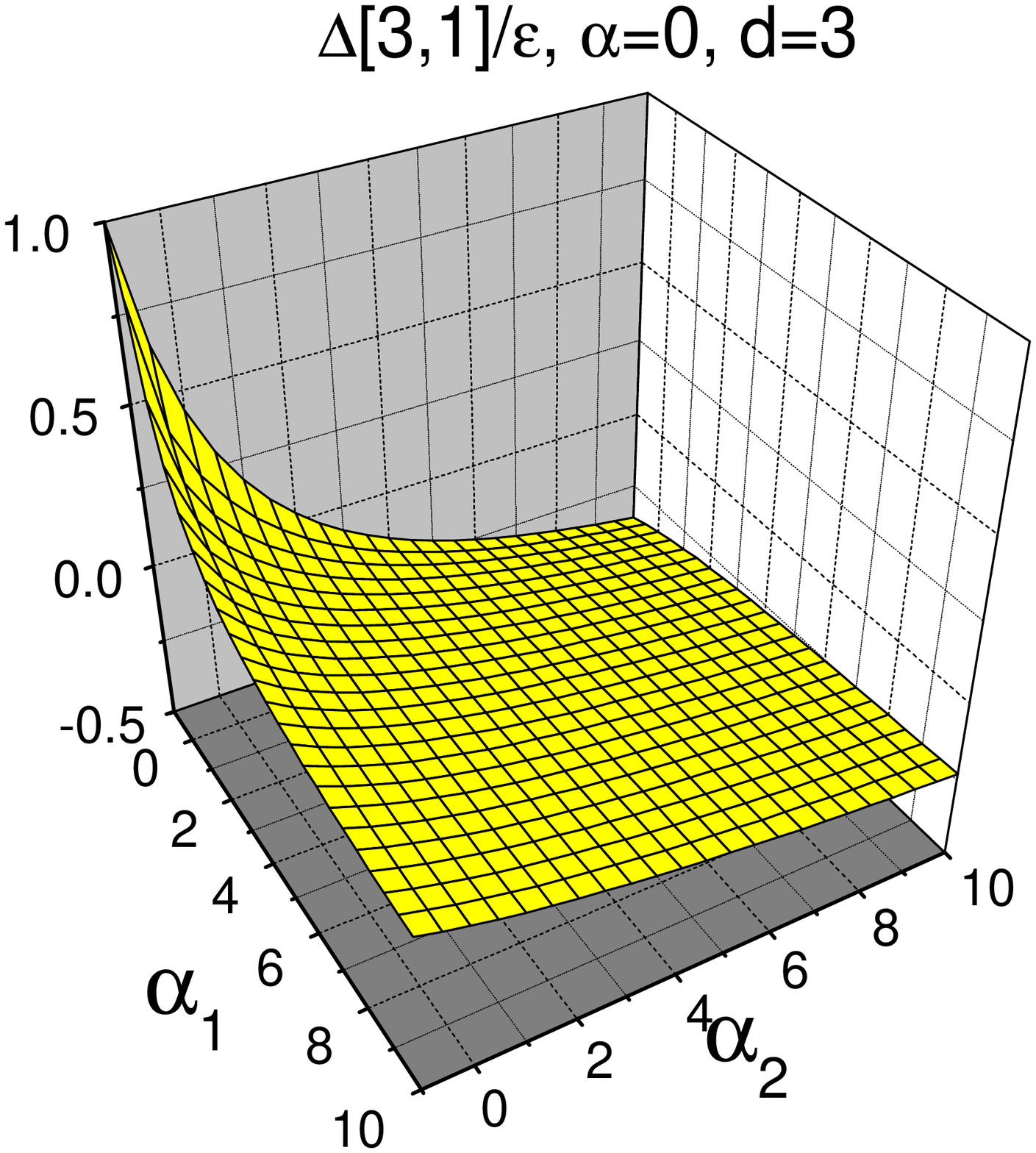}
   \end{flushleft}
     \vspace{-13.2cm}
   \begin{flushright}
       \leavevmode
       \epsfxsize=8.5cm
       \epsffile{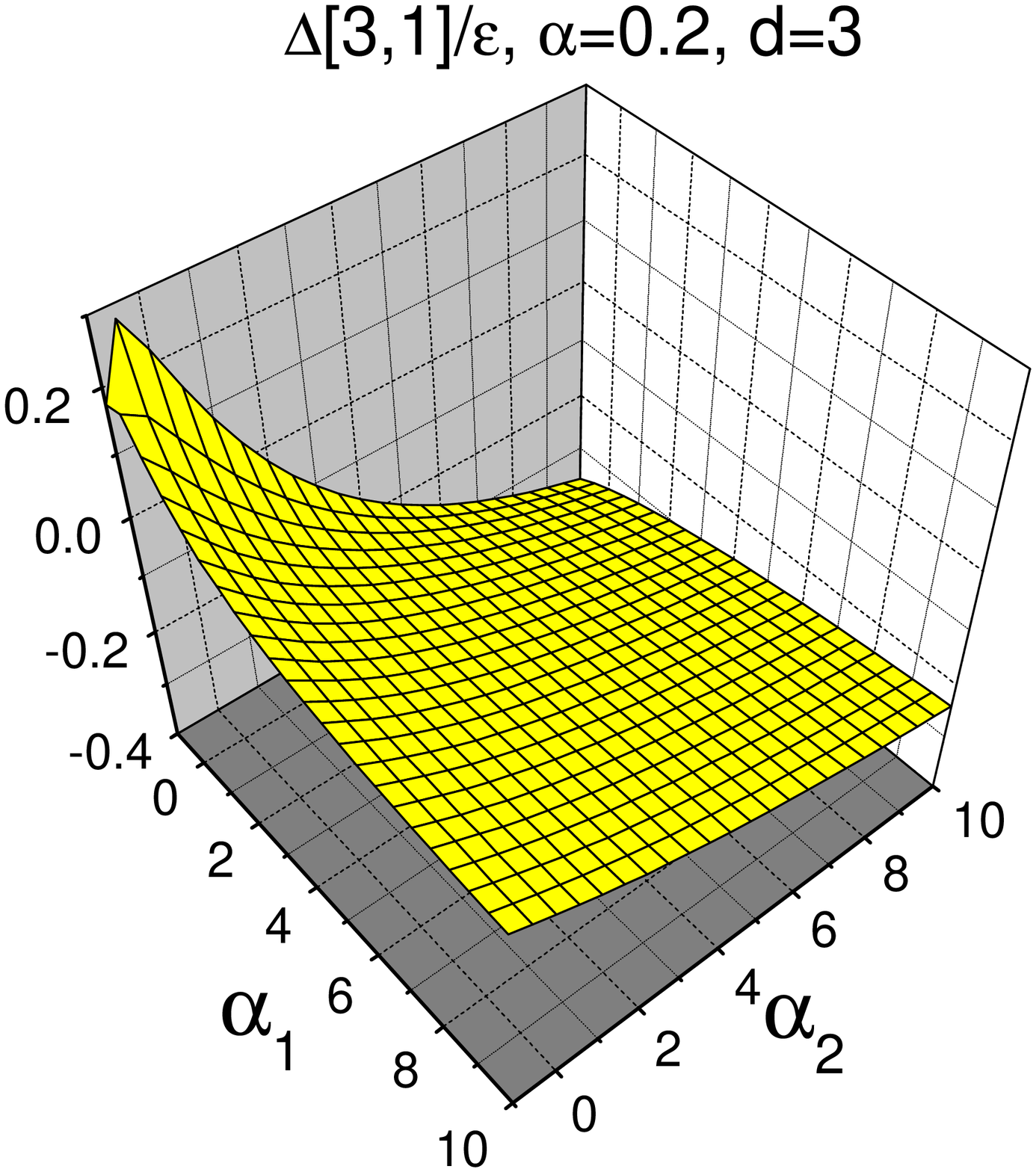}
   \end{flushright}
\vspace{-2.5cm} \caption{Behavior of the critical dimension
$\Delta[3,1]/\varepsilon$ on anisotropy parameter $\alpha_1$ and
$\alpha_2$ for $\alpha=0$ (incompressible case) and for
$\alpha=0.2$. \label{fig1112}}
\end{figure}

Let $F\equiv\{F_{\alpha}\}$ be a closed set of composite operators
which are mixed only with each other in renormalization. Then the
renormalization matrix $Z_F\equiv\{Z_{\alpha \beta}\}$ and the
matrix of corresponding anomalous dimensions $\gamma_F \equiv
\{\gamma_{\alpha\beta}\}$ for this set are given as follows
\begin{equation}
F_{\alpha}=\sum_{\beta}Z_{\alpha\beta}F_{\beta}^{R},
\qquad\gamma_{F}=Z_{F}^{-1} \tilde{D}_{\mu}Z_{F}.\label{2.2}
\end{equation}
Renormalized composite operators are subject to the following RG
differential equations
\begin{equation}
({\mathcal{{D}}}_{\mu}+\beta_{g}\partial_{g}+
\beta_{\chi}\partial_{\chi}-\gamma_{\nu}{\mathcal{{D}}}_{\nu})F_{\alpha}^{R}
=-\sum_{\beta}\gamma_{\alpha\beta}F_{\beta}^{R},
\end{equation}
which lead to the following matrix of critical dimensions
$\Delta_{F}\equiv\{\Delta_{\alpha\beta}\}$
\begin{equation}
\Delta_{F}=d_{F}^{k}-\Delta_{t}d_{F}^{\omega}+\gamma_{F}^{*},
\qquad\Delta_{t}=-2+2\varepsilon,\label{32B}
\end{equation}
where $d_{F}^{k}$ a $d_{F}^{\omega}$ are diagonal matrices of
corresponding canonical dimensions and $\gamma_{F}^{*}$ is the
matrix of anomalous dimensions (\ref{2.2}) taken at the fixed point.
In the end, the critical dimensions of the set of operators
$F\equiv\{ F_{\alpha}\}$ are given by the eigenvalues of the matrix
$\Delta_F$. The so-called "basis" operators that possess definite
critical dimensions have the form
\begin{equation}
F_{\alpha}^{bas}=\sum_{\beta}U_{\alpha\beta}F_{\beta}^{R}\,\,,\label{2.5}
\end{equation}
where the matrix $U_{F}=\{ U_{\alpha\beta}\}$ is such that
$\Delta'_{F}=U_{F}\Delta_{F}U_{F}^{-1}$ is diagonal.

\input epsf
   \begin{figure}[t]
     \vspace{-1cm}
       \begin{flushleft}
       \leavevmode
       \epsfxsize=8.5cm
       \epsffile{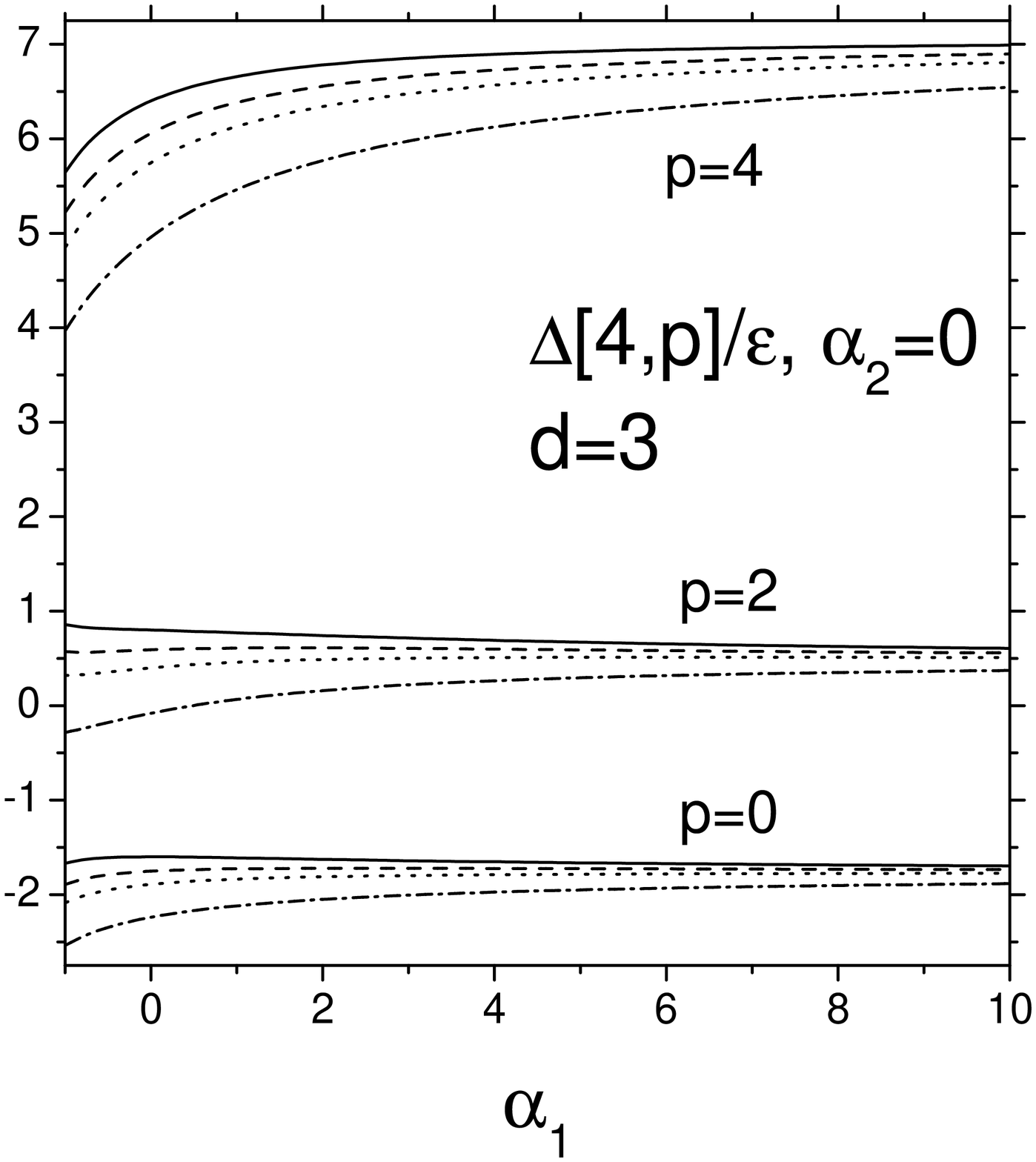}
   \end{flushleft}
     \vspace{-13.2cm}
   \begin{flushright}
       \leavevmode
       \epsfxsize=8.5cm
       \epsffile{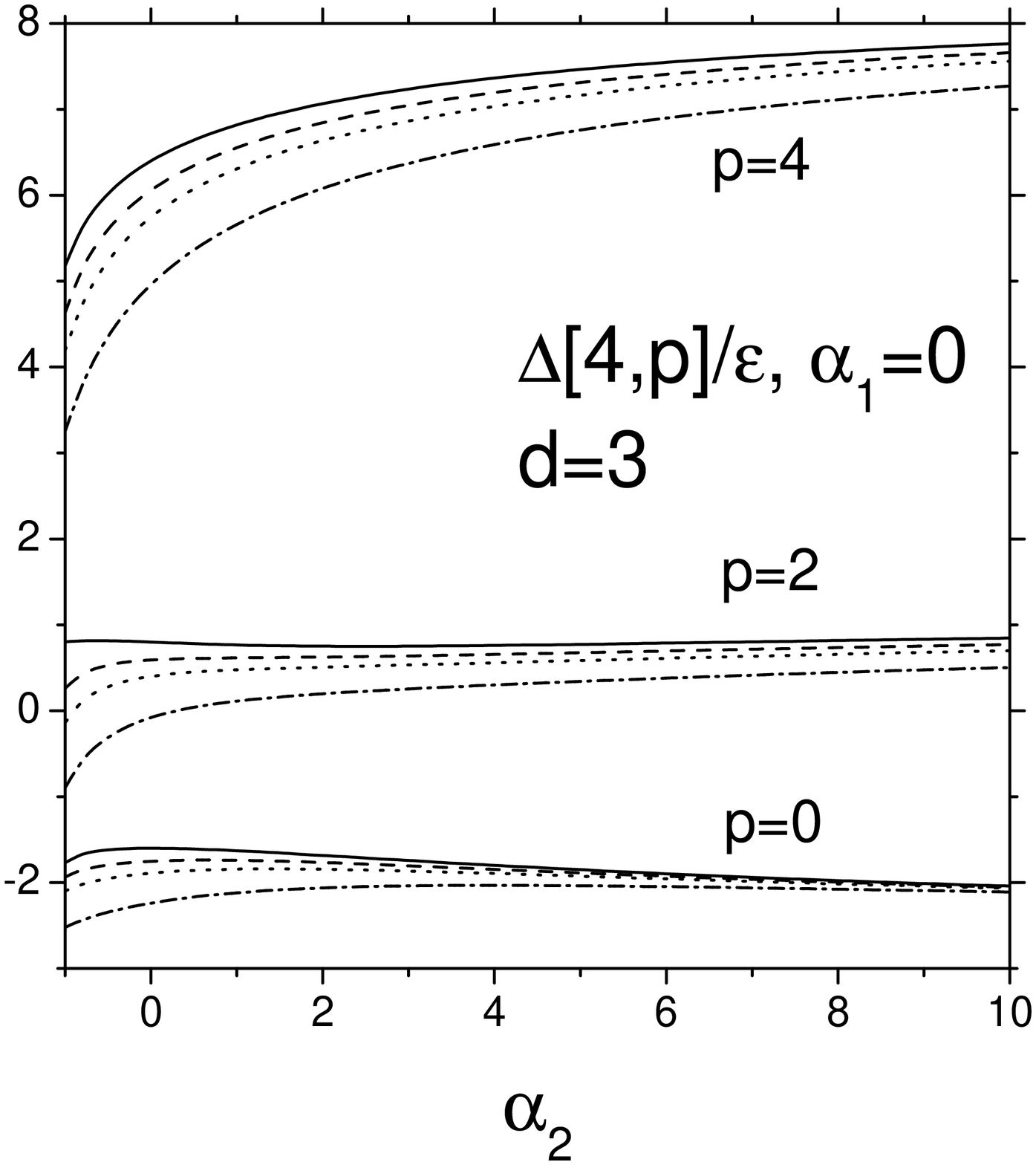}
   \end{flushright}
\vspace{-1.5cm} \caption{Dependence of the critical dimension
$\Delta[4,p]/\varepsilon, p=0,2,4$ on anisotropy parameters
$\alpha_1$ ($\alpha_2=0$) and $\alpha_2$ ($\alpha_1=0$) for
different values of the compressibility parameter $\alpha$ (solid
lines - $\alpha=0$, dash lines - $\alpha=0.1$, dot lines -
$\alpha=0.2$, and dash dot lines - $\alpha=0.5$). \label{fig1314}}
\end{figure}

\input epsf
   \begin{figure}[t]
     \vspace{-1cm}
       \begin{flushleft}
       \leavevmode
       \epsfxsize=8.5cm
       \epsffile{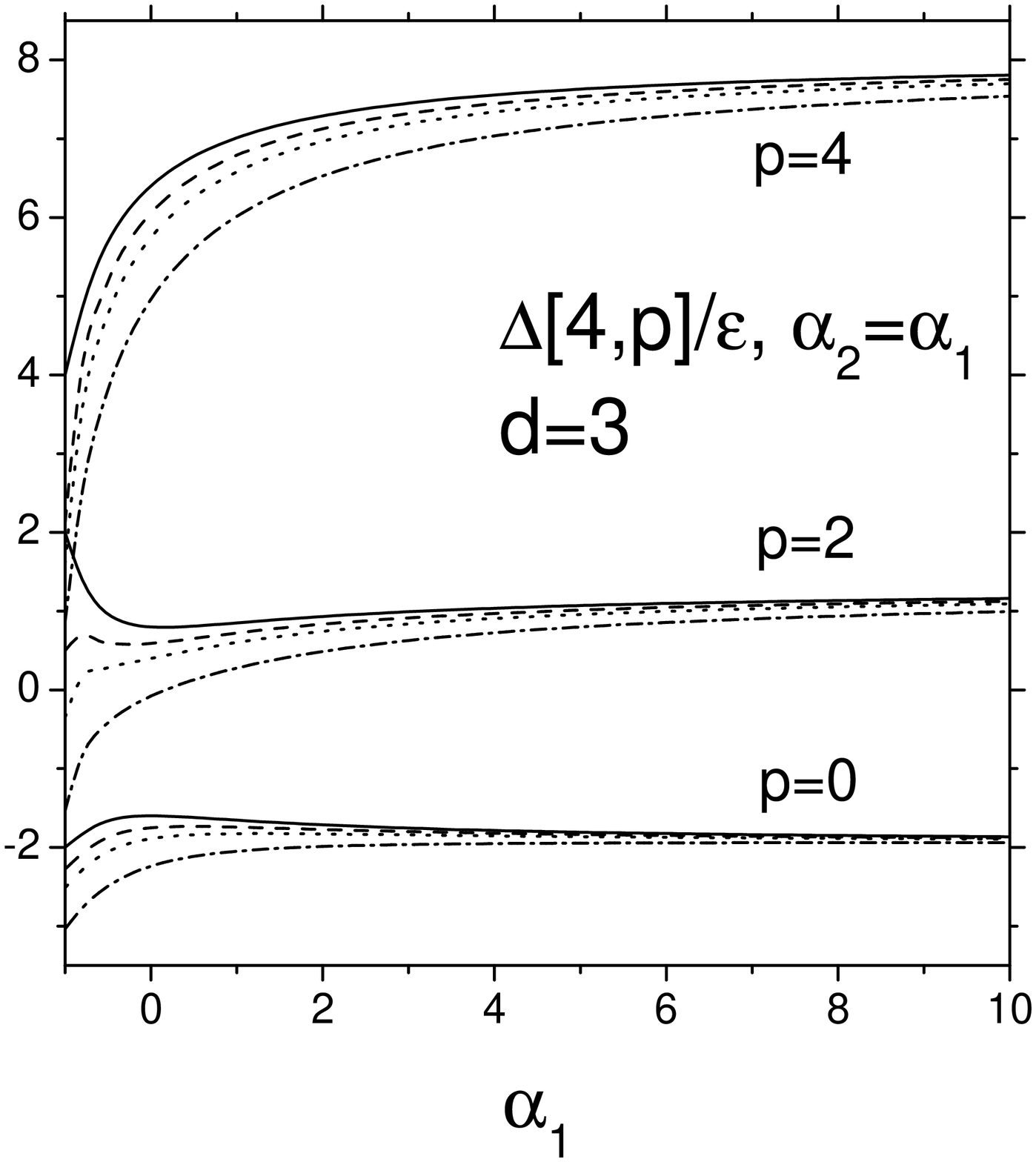}
   \end{flushleft}
     \vspace{-13.2cm}
   \begin{flushright}
       \leavevmode
       \epsfxsize=8.5cm
       \epsffile{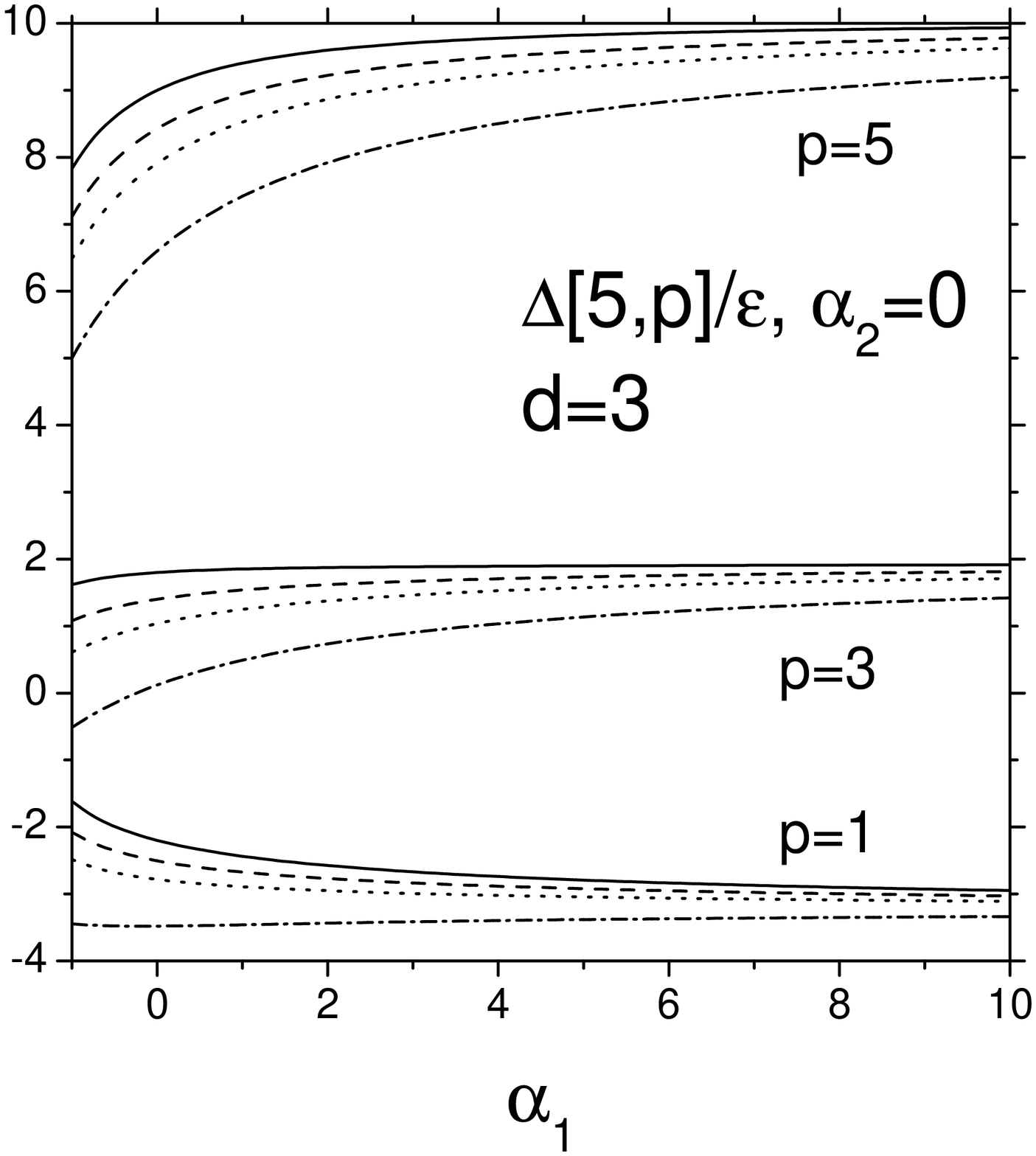}
   \end{flushright}
\vspace{-1.5cm} \caption{(Left) Dependence of the critical dimension
$\Delta[4,p]/\varepsilon, p=0,2,4$ on anisotropy parameters
$\alpha_1=\alpha_2$ for different values of the compressibility
parameter $\alpha$. (Right) Dependence of the critical dimension
$\Delta[5,p]/\varepsilon, p=1,3,5$ on anisotropy parameter
$\alpha_1$ ($\alpha_2=0$) for different values of the
compressibility parameter $\alpha$. See the caption in
Fig.\,\ref{fig1314} for line identification. \label{fig1516}}
\end{figure}

As was already mentioned, in our case of a scalar admixture with
anisotropy and compressibility the central role is played by the
tensor composite operators
$\partial_{i_{1}}\theta\cdots\partial_{i_{p}}\theta\,(\partial_{i}\theta\partial_{i}\theta)^{n}$,
constructed solely of the scalar gradients \cite{AdAnHnNo00}. It is
convenient to deal with the scalar operators obtained by contracting
the tensors with the appropriate number of the vectors ${\bf n}$,
namely
\begin{equation}
F[N,p]\equiv[(\mathbf{n\partial})\theta]^{p}(\partial_{i}\theta\partial_{i}\theta)^{n},\quad
N\equiv2n+p.\label{Fnp}
\end{equation}
Detail analysis shows that the composite operators (\ref{Fnp}) with
different $N$ are not mixed with each other in renormalization, that
is why the corresponding infinite renormalization matrix
\begin{equation}
F[N,p]=Z_{[N,p][N^{\prime},p^{\prime}]}F_R[N^{\prime},p^{\prime}]\,\label{renFk}
\end{equation}
is block-triangular, i.e., $Z_{[N,p][N^{\prime},p^{\prime}]}=0$ for
$N^{\prime}\neq N$. Thus, the critical dimensions associated with
the operator $F[N,p]$ are completely  determined by the eigenvalues
of the subblocks with $N^{\prime}=N$.

\input epsf
   \begin{figure}[t]
     \vspace{-1cm}
       \begin{flushleft}
       \leavevmode
       \epsfxsize=8.5cm
       \epsffile{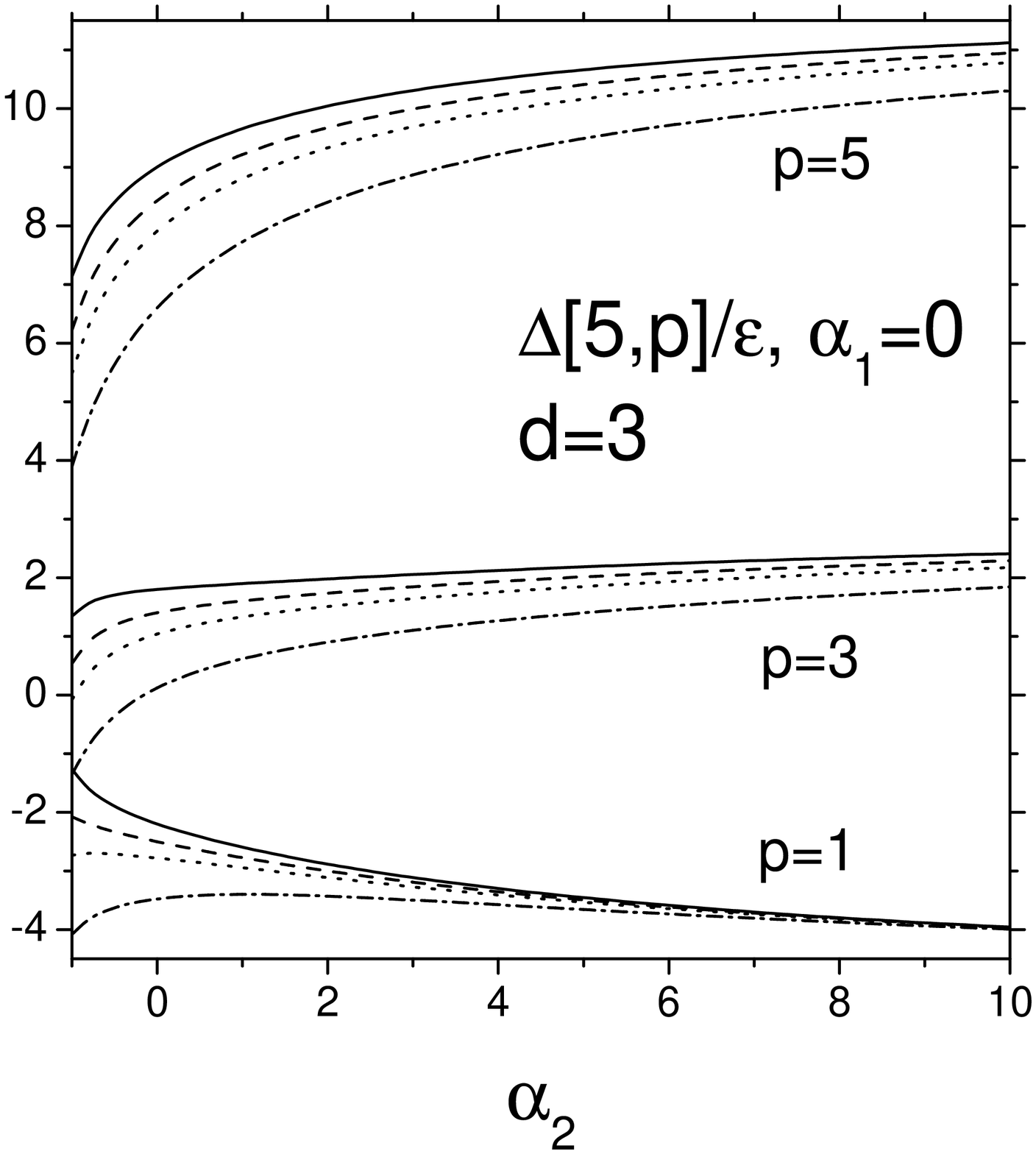}
   \end{flushleft}
     \vspace{-13.2cm}
   \begin{flushright}
       \leavevmode
       \epsfxsize=8.5cm
       \epsffile{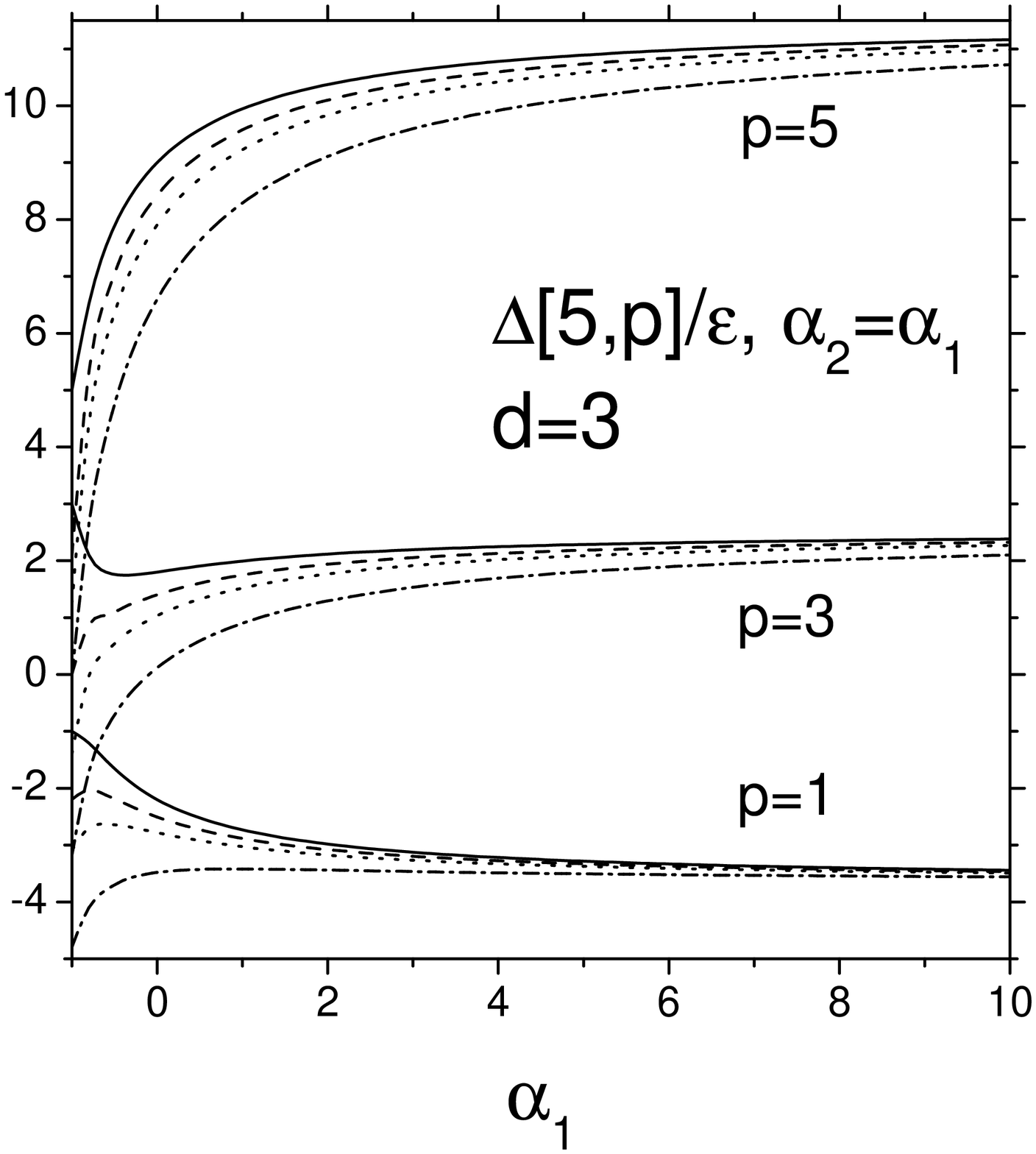}
   \end{flushright}
\vspace{-1.5cm} \caption{Dependence of the critical dimension
$\Delta[5,p]/\varepsilon, p=1,3,5$ on anisotropy parameters
$\alpha_2$ ($\alpha_1=0$) and $\alpha_1=\alpha_2$ for different
values of the compressibility parameter $\alpha$ (see the caption in
Fig.\,\ref{fig1314}).\label{fig1718}}
\end{figure}

In the isotropic case, as well as in the case when large-scale
anisotropy is present, the elements $Z_{[N,p]\,[N,p']}$ vanish for
$p<p'$, therefore the block $Z_{[N,p]\,[N,p']}$ is triangular. The
same is valid for the matrices $U_{F}$ and $\Delta_{F}$ defined in
Eqs.\,(\ref{2.5}) and (\ref{32B}). Thus, the presence of large-scale
anisotropy does not affect critical dimensions of the operators
(\ref{Fnp}). Situation is radically different in the case when
small-scale anisotropy is present, where the operators with
different values of $p$ mix in renormalization in such a way that
the matrix  $Z_{[N,p]\,[N,p']}$ is not triangular and one can write
\begin{equation}
F[N,p]=\sum_{l=0}^{\lfloor N/2 \rfloor} Z_{[N,p]\,[N,N-2l]}
F^R[N,N-2l]\,, \label{Fnl}
\end{equation}
where $\lfloor N/2 \rfloor$ means the integer part of the $N/2$.
Therefore, each block of renormalization constants with given $N$ is
an $(\lfloor N/2 \rfloor+1)\times(\lfloor N/2 \rfloor+1)$ matrix. Of
course, the matrix of critical dimensions (\ref{32B}), whose
eigenvalues at IR stable fixed point are the critical dimensions
$\Delta[N,p]$ of the set of operators $F[N,p]$, has also dimension
$(\lfloor N/2 \rfloor+1)\times(\lfloor N/2 \rfloor+1)$.

\input epsf
   \begin{figure}[t]
     \vspace{-1cm}
       \begin{flushleft}
       \leavevmode
       \epsfxsize=8.5cm
       \epsffile{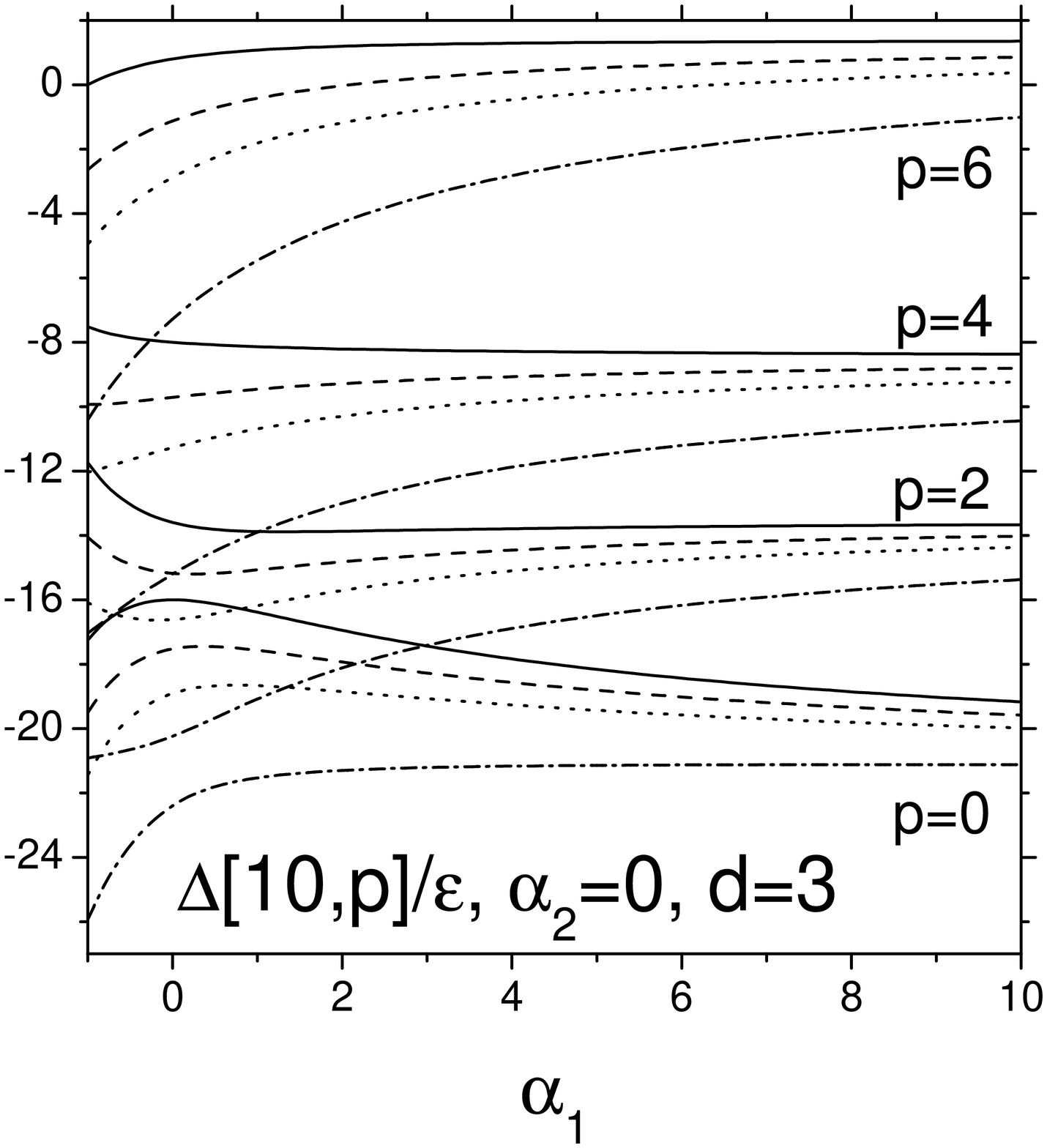}
   \end{flushleft}
     \vspace{-13.2cm}
   \begin{flushright}
       \leavevmode
       \epsfxsize=8.5cm
       \epsffile{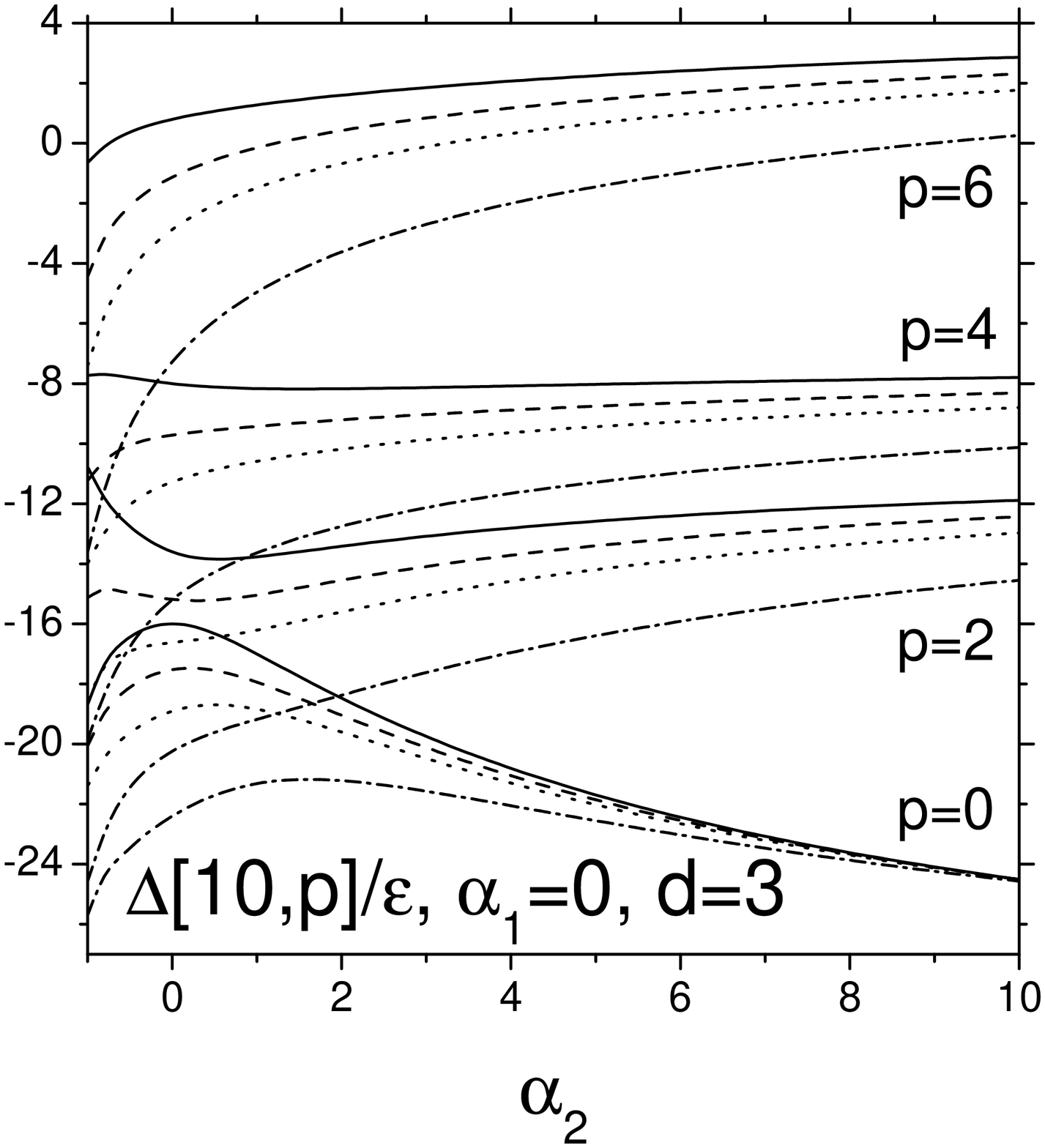}
   \end{flushright}
\vspace{-1.5cm} \caption{Dependence of the critical dimension
$\Delta[10,p]/\varepsilon, p=0,2,4,6$ on anisotropy parameter
$\alpha_1$ ($\alpha_2=0$) and $\alpha_2$ ($\alpha_1=0$) for
different values of the compressibility parameter $\alpha$ (see the
caption in Fig.\,\ref{fig1314}).\label{fig1920}}
\end{figure}

\input epsf
   \begin{figure}[t]
     \vspace{-1cm}
       \begin{flushleft}
       \leavevmode
       \epsfxsize=8.5cm
       \epsffile{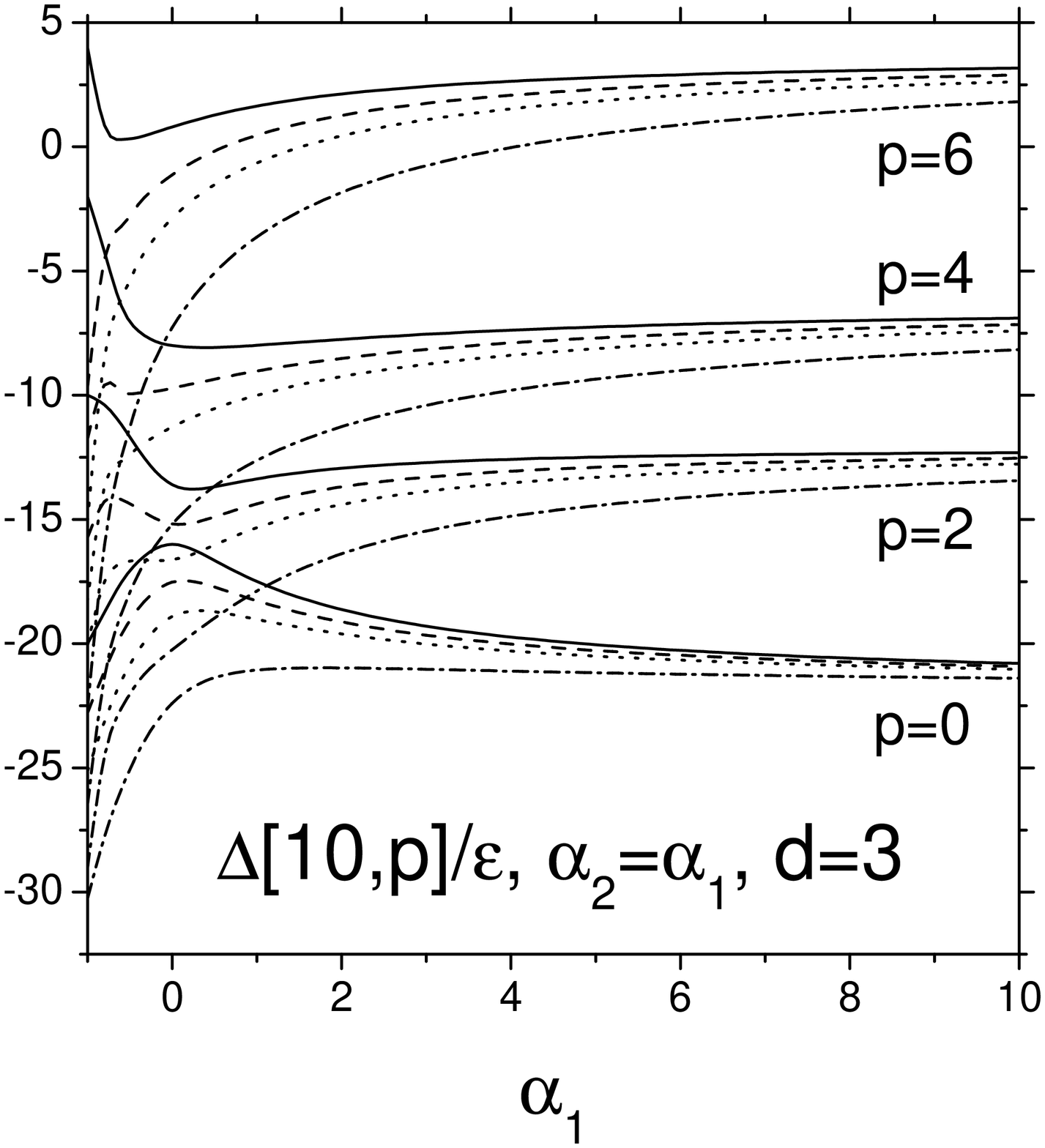}
   \end{flushleft}
     \vspace{-13.2cm}
   \begin{flushright}
       \leavevmode
       \epsfxsize=8.5cm
       \epsffile{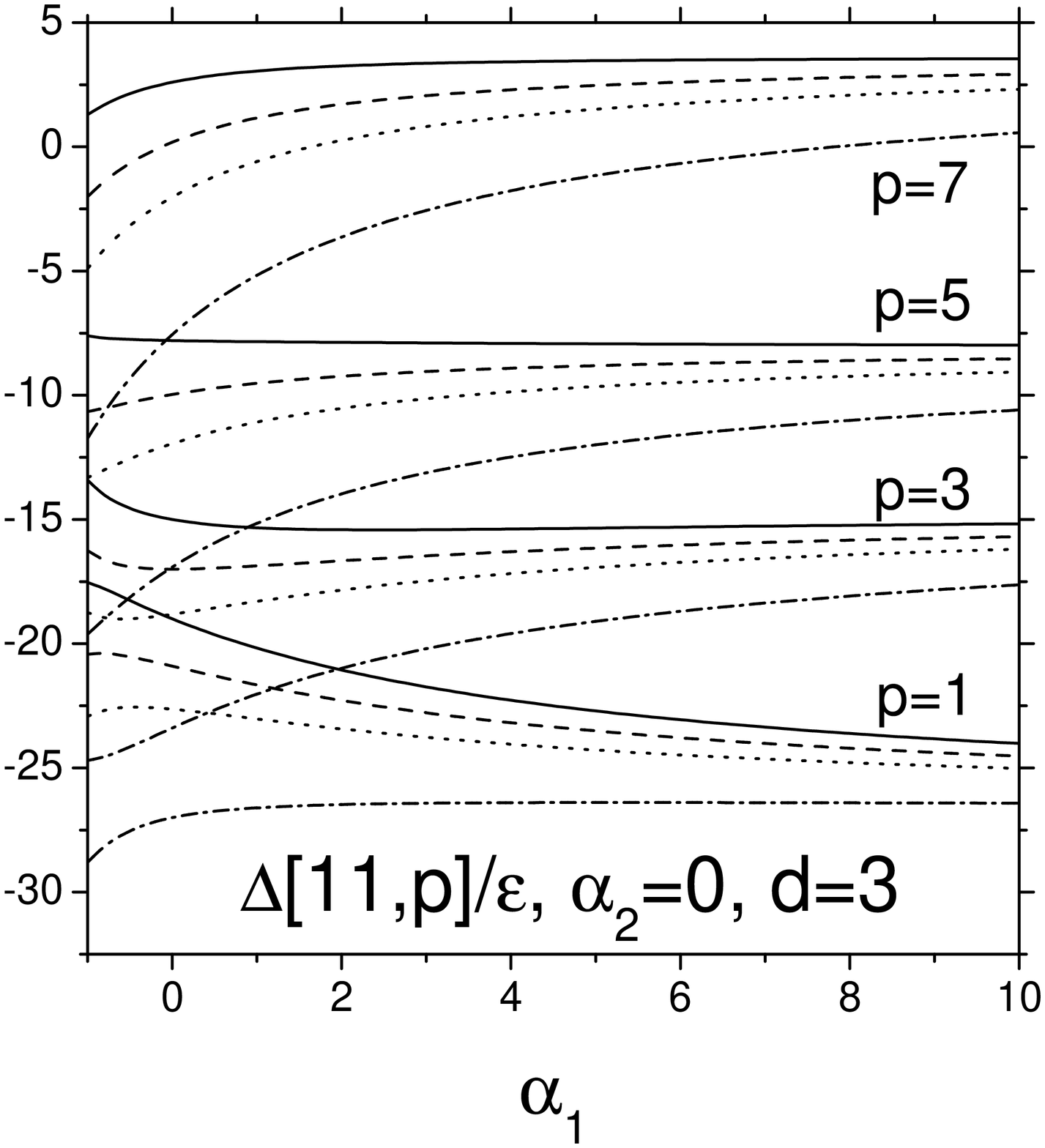}
   \end{flushright}
\vspace{-1.5cm} \caption{(Left) Dependence of the critical dimension
$\Delta[10,p]/\varepsilon, p=0,2,4,6$ on anisotropy parameters
$\alpha_1=\alpha_2$ for different values of the compressibility
parameter $\alpha$. (Right) Dependence of the critical dimension
$\Delta[11,p]/\varepsilon, p=1,3,5,7$ on anisotropy parameter
$\alpha_1$ ($\alpha_2=0$) for different values of the
compressibility parameter $\alpha$.  See the caption in
Fig.\,\ref{fig1314} for line identification. \label{fig2122}}
\end{figure}

In what follows, we shall calculate the matrix
$Z_{[N,p][N^{\prime},p^{\prime}]}$ in one-loop approximation (note
that in spite of the UV renormalization of our model, where one-loop
result is the complete solution of the problem because of
nonexistence of higher-loop corrections, the renormalization of the
composite operators has two- and higher-loop corrections). It was
analyzed in detail in Ref.\,\cite{AdAnHnNo00}, therefore we shall
not repeat it here, and we confine ourself only to the necessary
information.

We are interested in $N$-th term, denoted as $\Gamma_N[x;\theta]$,
of the expansion of the generating functional of
one-particle-irreducible Green functions with one composite operator
$F[N,p]$ and any number of fields $\theta$, denoted as
$\Gamma[x;\theta]$. It has the following form
\begin{equation}
\Gamma_{N}(x;\theta)=\frac{1}{N!}\int dx_{1}\cdots\int
dx_{N}\,\theta(x_{1})\cdots\theta(x_{N}) \langle
F[N,p](x)\theta(x_{1})\cdots\theta(x_{N})\rangle_{\textrm{1-ir}}.\label{Gamma1}
\end{equation}
In one-loop approximation the function (\ref{Gamma1}) is given as
\begin{equation}
\Gamma_{N}=F[N,p]+\Gamma^{(1)}\,, \label{Gamma2}
\end{equation}
where $\Gamma^{(1)}$ is given by the analytical calculation of the
diagram in Fig.\,\ref{fig4} (see, e.g., Ref.\,\cite{AdAnHnNo00} for
details) and the first term in Eq.\, (\ref{Gamma2}) represents
"tree" approximation.

After cumbersome but straightforward calculations (see
Ref.\,\cite{AdAnHnNo00}) one finds the UV divergent part of
$\Gamma^{(1)}$ from Eq.\,(\ref{Gamma2}) in the following appropriate
form
\begin{equation}
\Gamma^{(1)}= \frac{g S_d}{(2\pi)^d} \frac{1}{8 \chi^2 d^2
(d+1)(d+2)}\frac{1}{\varepsilon} \sum_{i=1}^{4} Q_i F[N,p+2(i-2)],
\label{oper1}
\end{equation}
with
\begin{equation}
Q_i=H_0(A_{i0}+\alpha A_{i1}) + H_1 (B_{i0} +\alpha
B_{i1})\,,\label{oper2}
\end{equation}
where $H_j=\, _{2}F_{1}\left(\frac{1}{2}, 1;
j+\frac{d}{2};-\chi\right), j=0,1$ are corresponding hypergeometric
functions, and coefficients $A_{ij}$ and $B_{ij}$ for $i=1,2,3,4$
and $j=0,1$ are given in Appendix A.

\input epsf
   \begin{figure}[t]
     \vspace{-1cm}
       \begin{flushleft}
       \leavevmode
       \epsfxsize=8.5cm
       \epsffile{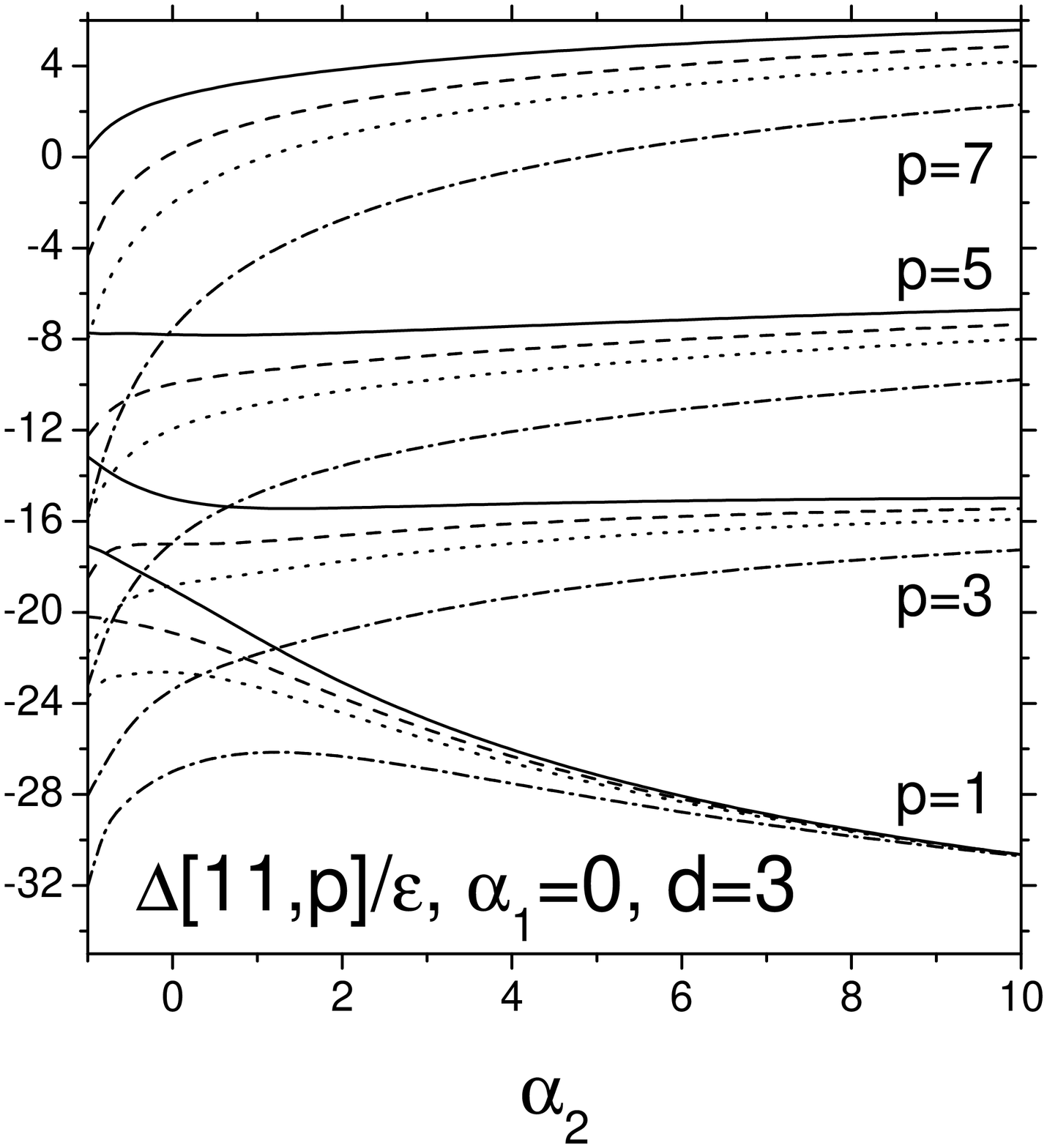}
   \end{flushleft}
     \vspace{-13.2cm}
   \begin{flushright}
       \leavevmode
       \epsfxsize=8.5cm
       \epsffile{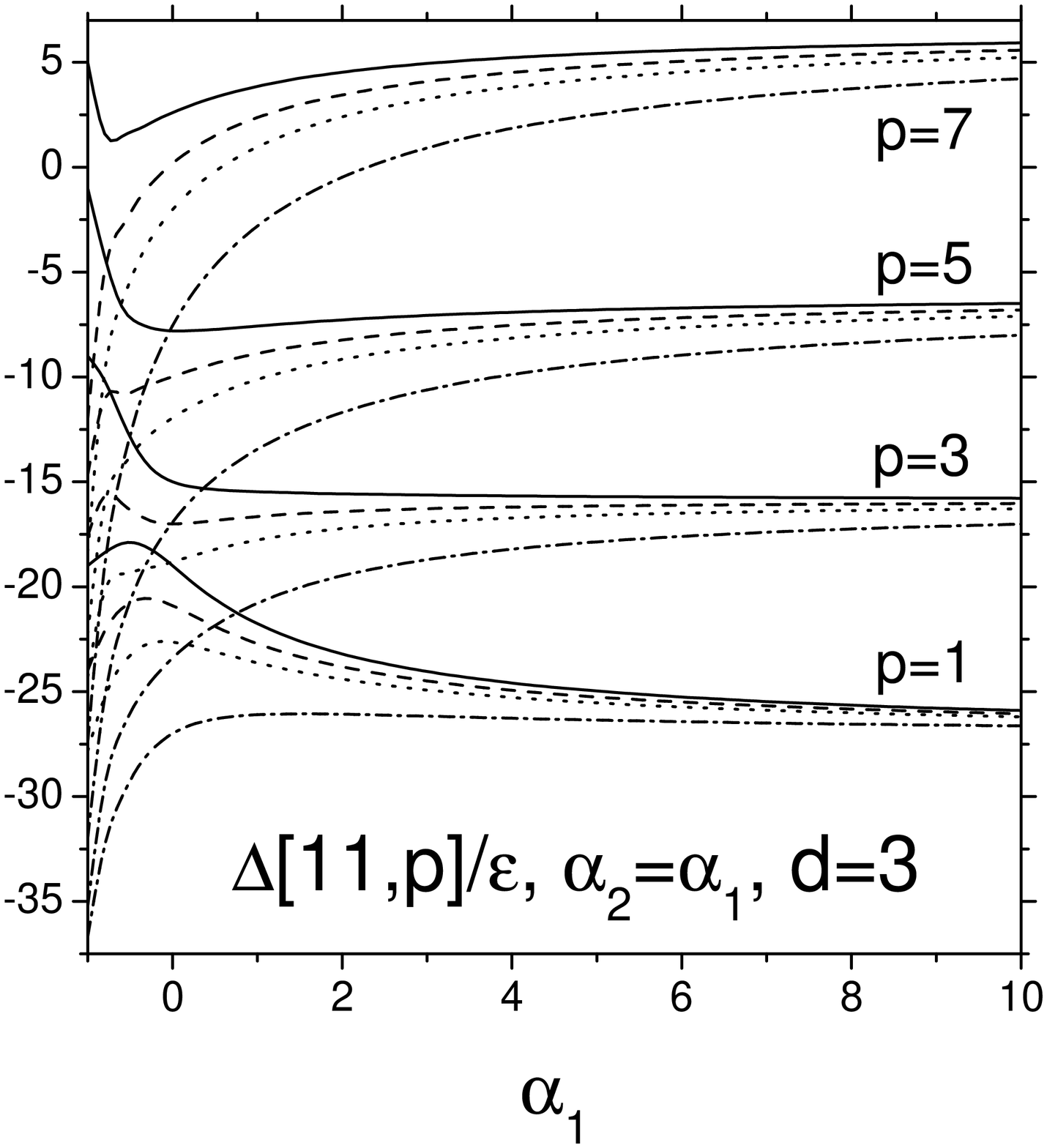}
   \end{flushright}
\vspace{-1.5cm} \caption{Dependence of the critical dimension
$\Delta[11,p]/\varepsilon, p=1,3,5,7$ on anisotropy parameter
$\alpha_2$ ($\alpha_1=0$) and $\alpha_1=\alpha_2$ for different
values of the compressibility parameter $\alpha$ (see the caption in
Fig.\,\ref{fig1314}). \label{fig2324}}
\end{figure}

\input epsf
   \begin{figure}[t]
     \vspace{-1cm}
       \begin{flushleft}
       \leavevmode
       \epsfxsize=8.5cm
       \epsffile{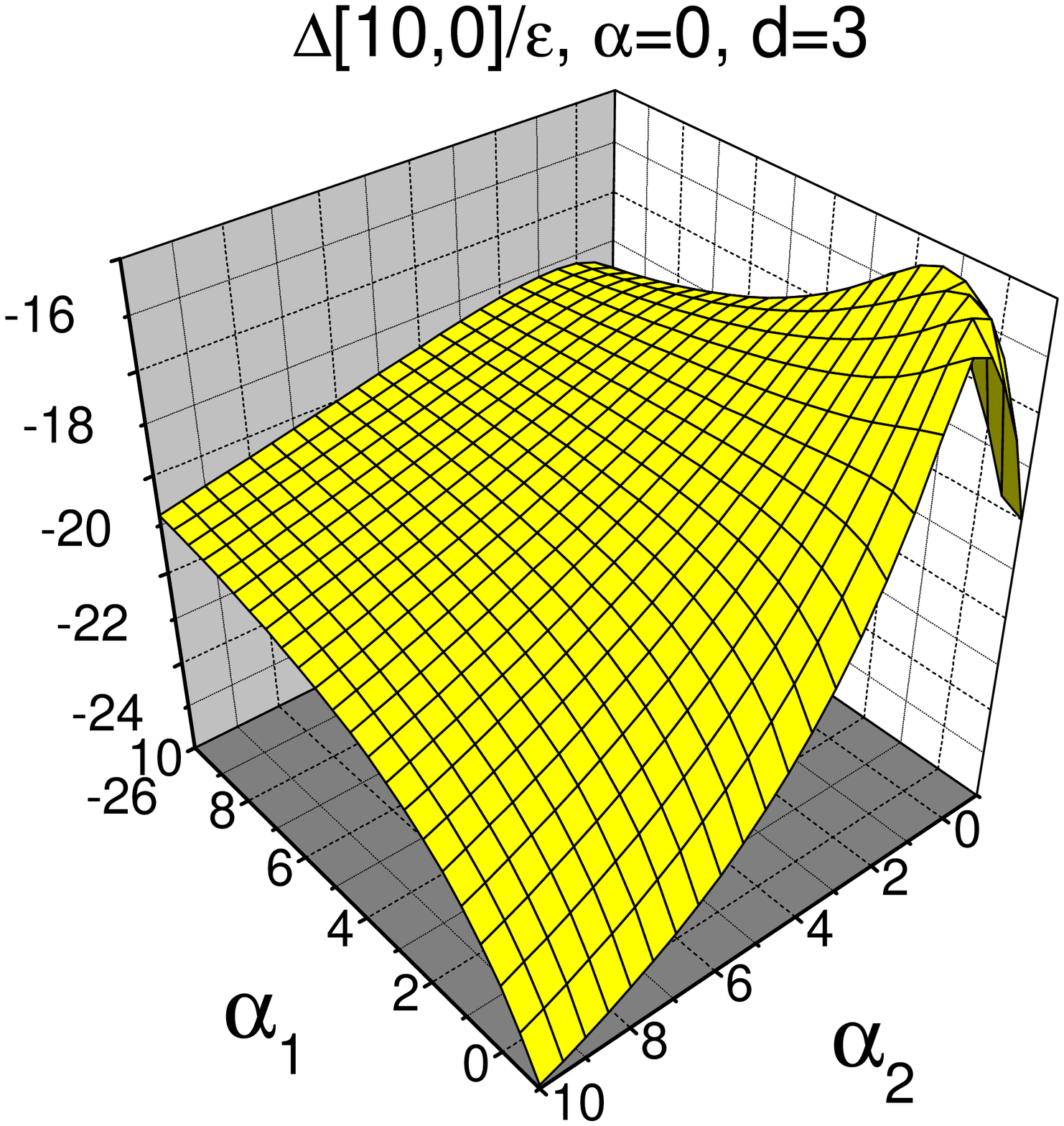}
   \end{flushleft}
     \vspace{-13.2cm}
   \begin{flushright}
       \leavevmode
       \epsfxsize=8.5cm
       \epsffile{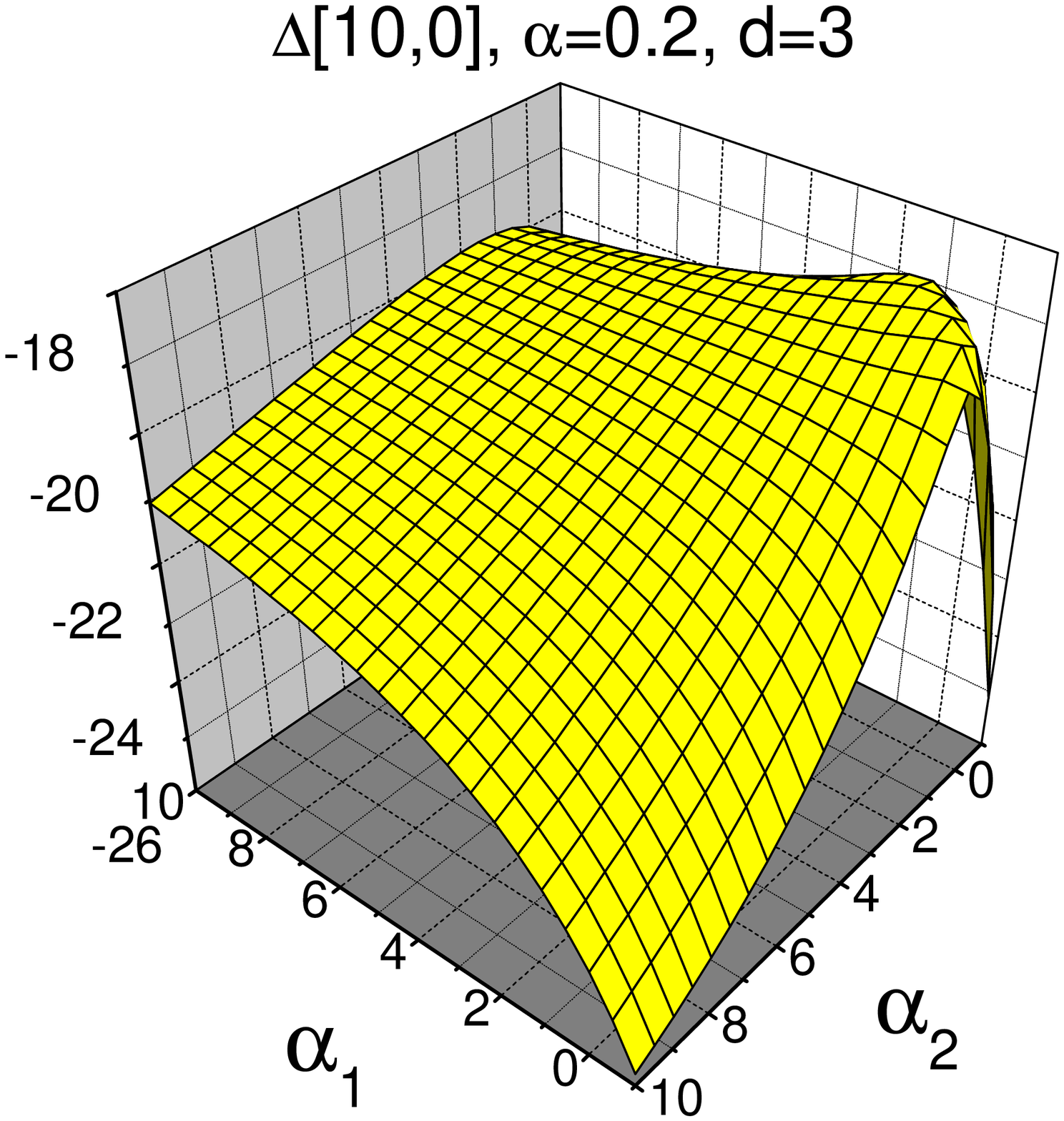}
   \end{flushright}
\vspace{-2.5cm} \caption{Behavior of the critical dimension
$\Delta[10,0]/\varepsilon$ on anisotropy parameters $\alpha_1$ and
$\alpha_2$ for $\alpha=0$ (incompressible case) and for
$\alpha=0.2$. \label{fig2526}}
\end{figure}

The renormalization matrix $Z_{[N,p]\,[N,p']}$ is then found from
the condition that  $F^{R}[N,p]$ from Eq.\,(\ref{Gamma2}) must be
free of UV divergences when are written in renormalized variables,
i.e., it does not contain poles in $\varepsilon$.  In the MS scheme
one obtains
\begin{equation}
Z_{[N,p][N,p+2(i-2)]} = \delta_{2i}+\frac{g\,
S_{d}}{(2\pi)^d}\frac{A}{8\varepsilon}\, Q_{i} \label{Znp}
\end{equation}
where $i=1,2,3,4$, $A=1/(d^2\chi^2(d+2)(d+1))$, and coefficients
$Q_{i}, i=1,2,3,4$ are given in  Eq.\,(\ref{oper2}). In the end,
from Eq.\,(\ref{Znp}) and using definitions (\ref{2.2}), one finds
the following expressions for the elements of the matrix of
anomalous dimensions $\gamma_{[N,p]\,[N',p']}$
\begin{equation}
\gamma_{[N,p][N,p+2(i-2)]} = -\frac{g\,
S_{d}}{(2\pi)^d}\frac{A}{4}\, Q_{i}
\end{equation}
for $i=1,2,3,4$ and  the matrix of critical dimensions (\ref{32B})
has the form
\begin{equation}
\Delta_{[N,p][N,p']}=N\,\varepsilon+\gamma_{[N,p][N,p']}^{*},\label{Dnp}
\end{equation}
where the asterisk means that $\gamma_{[N,p][N,p']}$ is taken at the
fixed point given by Eqs.\,(\ref{gAA1}) and (\ref{hAA2}).
Eq.\,(\ref{Dnp}), which depends on the anisotropy parameters
$\alpha_1,\alpha_2$, as well as on the compressibility parameter
$\alpha$, is desired one-loop expression for the matrix of critical
dimensions of the composite operators (\ref{Fnp}).

\subsection{ Anomalous scaling: one-loop
approximation}

The critical dimensions of the operators $F[N,p]$, which we denote
as $\Delta[N,p]$, are after all equal to the eigenvalues of the
matrix of critical dimensions (\ref{Dnp}). In the isotropic case or
in the case with large-scale anisotropy, where the matrix
(\ref{Dnp}) is triangular, the eigenvalues are equal directly to the
diagonal elements of the matrix. This fact allows us to assign
uniquely the concrete critical dimension to the corresponding
composite operator even in the case with small-scale anisotropy and
study their hierarchical structure as a functions of $p$ (see
\cite{AdAnHnNo00} for details). In Ref.\,\cite{AdAnHnNo00} it was
shown that some of the critical dimensions are negative, therefore
they lead to the anomalous scaling (singular behavior of the scaling
functions). At the same time, the leading role is played by the
operators with the most negative critical dimensions which are the
operators with $p=0$ for the structure functions (\ref{struc}) with
even $N$, and the operators with $p=1$ for the structure functions
with odd $N$.

Thus, the combination of the RG representation with the OPE
(\ref{ope}) leads to the final asymptotic expression for the
structure functions (\ref{struc}) within the inertial range
\begin{equation}
S_N({\bf r})=D_0^{-N/2} r^{N(1-\varepsilon)} \sum_{N^{\prime}\leq N}
\sum_p
\{C_{N^{\prime,p}}\,(r/L)^{\Delta[N^{\prime},p]}+\dots\}\,,\label{struc10}
\end{equation}
where $p$ obtains all possible values for given $N^{\prime}$,
$C_{N^{\prime,p}}$  are numerical coefficients which are functions
of the parameters of the model, and dots means contributions by the
operators others than $F[N,p]$ (see, e.g.,
\cite{Vasiliev,AdAnHnNo00} for details).

In what follows, we shall investigate the influence of
compressibility on the picture found in Ref.\,\cite{AdAnHnNo00}. Our
aim is to find the answer on the question whether the presence of
compressibility leads to the more pronounced anomalous scaling of
structure functions of the scalar field or not. Another purpose of
the present analysis is to check whether the combine effects defined
by compressibility and small-scale anisotropy can lead to the more
complicated structure of critical dimensions than it was studied in
Ref.\,\cite{AdAnHnNo00} (see also Ref.\,\cite{HnHoJuMaSp05}). One
possibility is that the pairs of complex conjugate eigenvalues of
the matrix of critical dimensions can exist. It leads to the
oscillation behavior of the corresponding scaling function
\cite{AdAnHnNo00}, i.e., the scaling functions in
Eq.\,.(\ref{struc10}) would contain terms of the following form
\begin{equation}
(r/L)^{\Delta_{R}}\left\{ c_1 \cos\left[\Delta_I (r/L) \right] + c_2
\sin\left[\Delta_I (r/L) \right]\right\},
\end{equation}
where $\Delta_R$ and $\Delta_I$ are real and imaginary part of
$\Delta$, and $c_{1,2}$ are constants. Another possibility is
related to the situation if the matrix of critical dimensions cannot
be diagonalized and has only the Jordan form. Then a logarithmic
correction would be involved to the powerlike behavior (see
Ref.\,\cite{AdAnHnNo00}).

\input epsf
   \begin{figure}[t]
     \vspace{-1cm}
       \begin{flushleft}
       \leavevmode
       \epsfxsize=8.5cm
       \epsffile{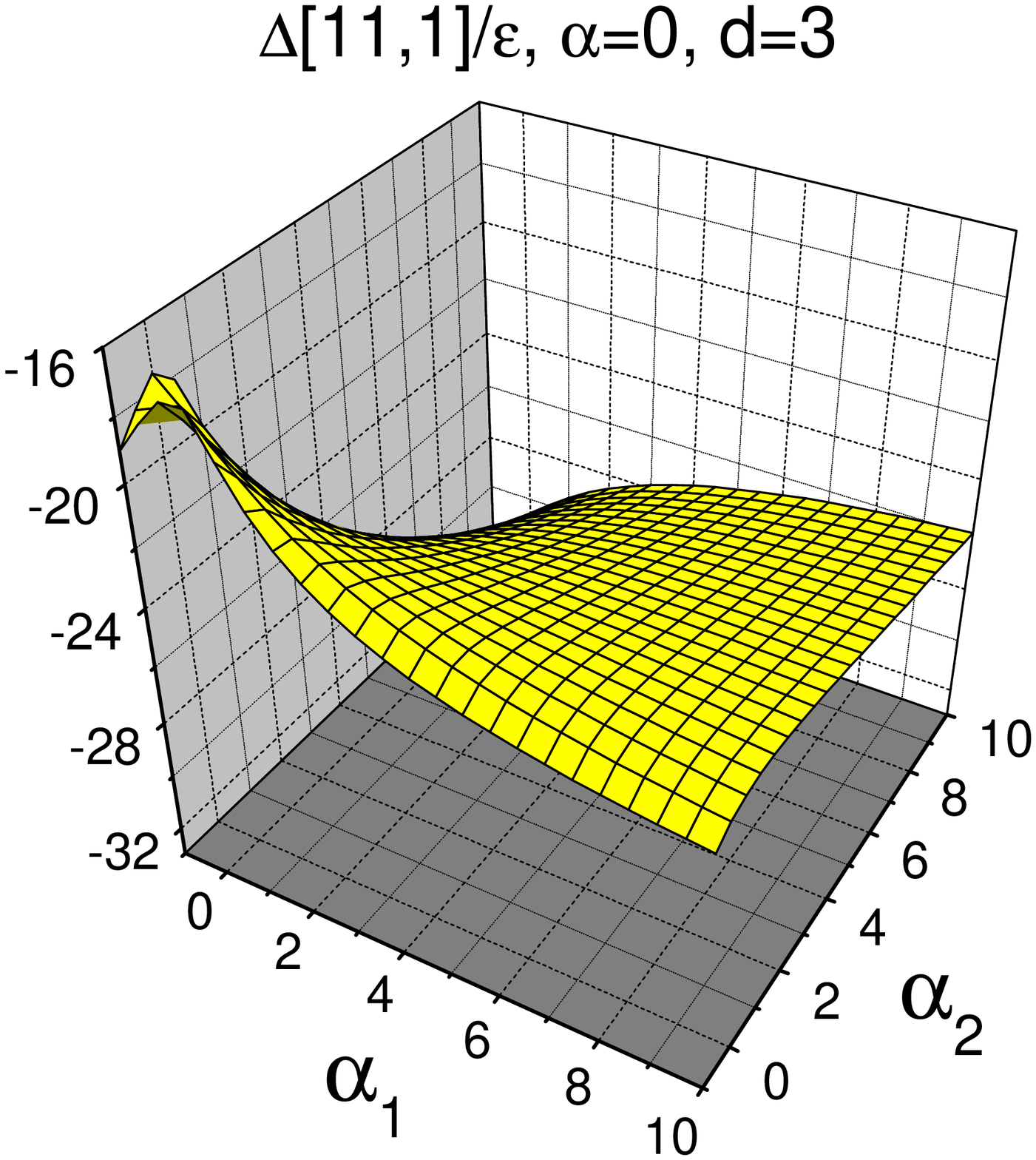}
   \end{flushleft}
     \vspace{-13.2cm}
   \begin{flushright}
       \leavevmode
       \epsfxsize=8.5cm
       \epsffile{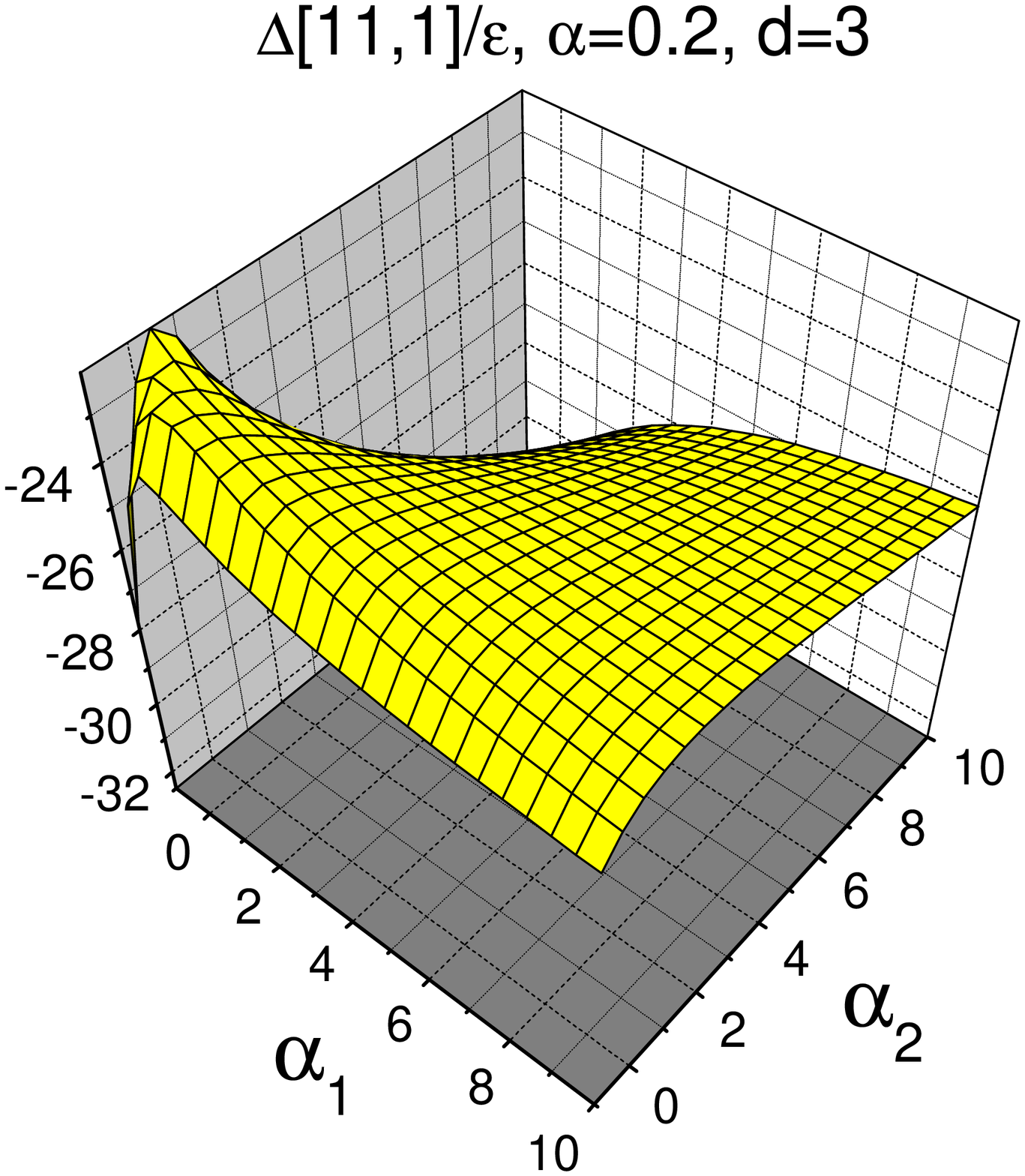}
   \end{flushright}
\vspace{-2.5cm} \caption{Behavior of the critical dimension
$\Delta[11,1]/\varepsilon$ on anisotropy parameters $\alpha_1$ and
$\alpha_2$ for $\alpha=0$ (incompressible case) and for
$\alpha=0.2$. \label{fig2728}}
\end{figure}

Behavior of the eigenvalues of the matrix of critical dimensions
$\Delta[N,p]$ for various values of $N$ for different values of the
compressibility parameter $\alpha$ and as the functions of the
anisotropy parameters $\alpha_1$ and $\alpha_2$ are shown in
Figs.\,\ref{fig56}-\ref{fig2728}. It can be immediately seen that
only real eigenvalues of the corresponding matrix exist as in the
incompressible case studied in Ref.\,\cite{AdAnHnNo00}. At the same
time, the hierarchical structure of the critical dimensions for
different values of $p$ and the same $N$ shown in
Ref.\,\cite{AdAnHnNo00} is also preserved. The dependence of the
critical dimension $\Delta[2,0]$ is not shown because it is
identically equal to zero. It can be shown either by direct
calculation or by using the Schwinger equation (see, e.g.,
Ref.\,\cite{AdAnHnNo00}). All the figures show that compressibility
leads to smaller values of the critical dimensions of the
corresponding composite operators, thus, at the same time, to the
more noticeable anomalous scaling. As it is shown in the figures,
the effects of compressibility are the most pronounced for small
values of the anisotropy parameters, as well as for the negative
values of these parameters. On the other hand, in the limit of large
positive values of anisotropy parameters $\alpha_1$ and $\alpha_2$
the effects of compressibility tend to zero. Especially, in the
limit $\alpha_{1,2}\rightarrow \infty$ the influence of
compressibility is vanished completely which can be shown by direct
analytical investigation.

\section{Conclusion}\label{sec:Conc}

In this paper we have analyzed asymptotic behavior of the
single-time structure functions $S_N(r)$ of a passively advected
scalar field by the compressible velocity field with small-scale
anisotropy by using the field-theoretic renormalization group and
the operator product expansion in a minimal-subtraction scheme of
analytical renormalization.

It is shown that the leading-order powerlike asymptotic behavior of
the single-time structure functions of a scalar field within the
inertial range in our compressible anisotropic case is given by the
critical dimensions of the same composite operators as in the
incompressible anisotropic case but now they acquire rather strong
dependence on the compressibility parameter. Further, it is shown
that when the parameter of compressibility is increasing then the
critical dimensions of relevant composite operators become smaller,
therefore, within our model, we can conclude that the anomalous
scaling of the structure functions of a passive scalar quantity
advected by a given velocity field is more pronounced in the
compressible stochastic environment than in the incompressible one
(see Figs.\,\ref{fig56}-\ref{fig2728}). Concrete calculations were
done up to the structure functions of order $N=11$. In our
calculations we have not found possible oscillatory modulation
(related to the possible existence of the complex conjugate
eigenvalues of the matrix of critical dimensions), as well as we
have not found possible logarithmic  corrections (related to the
fact that the corresponding matrix can has only Jordan form) to the
leading powerlike asymptotic. Thus, all calculated corrections have
had purely powerlike behavior.

It is also shown that the critical dimensions are ordered
hierarchically as in the incompressible case \cite{AdAnHnNo00},
i.e., the compressibility does not disturb the anisotropic structure
of the critical dimensions rather it shifts them toward smaller
values. The largest shift of the critical dimensions is present for
negative and small positive values of the anisotropy parameters. On
the other hand, when the anisotropy parameters tend to infinity the
effects of compressibility vanish.

\vspace{0.8cm} \noindent {\bf Acknowledgments} \vspace{0.2cm}

\noindent The work was supported in part by VEGA grant 6193 of
Slovak Academy of Sciences and by Science and Technology Assistance
Agency under contract No. APVT-51-027904.

\vspace{0.8cm} \noindent {\bf Appendix} \vspace{0.2cm}

The explicit form of the coefficients $A_{ij}$ and $B_{ij}$ with
$i=1,2,3,4$ and $j=0,1$ (see Eq.\,(\ref{oper2})) is
\begin{eqnarray}
A_{10}&=& p (p-1) d (1 + d) ((\alpha_1 - \alpha_2) (2 + d)  - (d+2 +
\alpha_2(1 + d) + \alpha_1 (d^2-3)) \chi  \nonumber
\\ && +(d-2 + \alpha_2 + \alpha_1(1 + d) + d^2) \chi^2)),\nonumber
\end{eqnarray}
\begin{eqnarray}
A_{11}&=& p (p-1) d (1 + d) (2 + d) (1 + \chi) \chi,\nonumber
\end{eqnarray}
\begin{eqnarray}
B_{10}&=& p (p-1) (-(\alpha_1 - \alpha_2) d (1 + d) (2 + d)
+ (1 + d) (2 + d) (\alpha_2 (d-1) + \alpha_1 (d-1)^2 + d) \chi \nonumber \\
&& - (d^2-1) (-2 + \alpha_1 + \alpha_2 + d + \alpha_1 d + d^2)
\chi^2), \nonumber
\end{eqnarray}
\begin{eqnarray}
B_{11}&=& p (p-1) (-(1 + d) (2 + d) d \chi  - (d+2) (d^2-1)
\chi^2),\nonumber
\end{eqnarray}
\begin{eqnarray}
A_{20}&=&d (4 n(n-1) (1 + \chi) (3 (2 + d)( \chi-\alpha_1) +
\alpha_1 (d^2-1) \chi \nonumber
\\ &&  + 3 \alpha_2 (2 + d + \chi(1+d))) - (1 + d) (p(p-1) (-d (2 + d) \chi
+ \alpha_1 (d (2 + d)\nonumber \\
&& + 2 \chi (1 + \chi)) - \alpha_2 (1 + \chi) (d (2 + d(1 + \chi))-2
\chi)) \nonumber \\ &&- 2 ((2 + d) \chi (\alpha_2-\alpha_1 +
(\alpha_2 + \alpha_1 (d-1) + d) \chi) \nonumber \\ &&- 2 (-(2 + d)
\chi ((d-3) \chi-3) + \alpha_2 (1 + \chi) (6 + 3 d + \chi + 2 d
\chi) \nonumber \\ && + \alpha_1 (d^2 \chi + (\chi-6) (1 + \chi) - d
(3 + \chi (2 + \chi)))) p) n)),\nonumber
\end{eqnarray}
\begin{eqnarray}
A_{21} &=& d (-12 n(n-1) (1 + \chi) (2 + d) \chi  + (1 + d) (-p(p-1) d (2 + d) \chi  \nonumber \\
&& + 12 (1 + \chi) (2 + d) \chi p n)), \nonumber
\end{eqnarray}
\begin{eqnarray}
B_{20} &=& -4 n(n-1) (-3 (\alpha_1 - \alpha_2) d (2 + d)  + (2 + d)
(3 d -3 \alpha_2 + 6 \alpha_2 d \nonumber
\\ && + \alpha_1 (3 + (d-5) d)) \chi + (7
d -6  - 3 \alpha_2 + (3 + 3 \alpha_2 - d) d^2 \nonumber
\\ && + \alpha_1 (d-1)^2 (1
+ d)) \chi^2) + (1 + d) (p(p-1) ((\alpha_1 - \alpha_2) d^2 (2 + d)
\nonumber \\ && - d (2 + d) (2 \alpha_2 (-1 + d) + d) \chi  - (d-1)
(-2 (\alpha_1 + \alpha_2) + \alpha_2 d^2) \chi^2) \nonumber \\ &&- 2
((2 + d) \chi ((\alpha_2 - \alpha_1) d + (\alpha_2 (d-1) + \alpha_1
(d-1)^2 + d) \chi) \nonumber
\\ && + 2 (3 (\alpha_1 - \alpha_2) d (2 + d)
- (2 + d) (\alpha_1 (d-3) (d-1) + 3 d -3 \alpha_2 + 5 \alpha_2 d) \chi \nonumber \\
&& + (d-1) (-6 - \alpha_1 - \alpha_2 + (\alpha_1 - 2 \alpha_2-1) d +
d^2) \chi^2) p) n),\nonumber
\end{eqnarray}
\begin{eqnarray}
B_{21} &=& -12 n(n-1) (-(2 + d) d \chi + (2 - 3 d - 2 d^2) \chi^2) -
(1 + d) (-p(p-1) d^2 (2 + d)  \chi \nonumber \\ && + 2 (-(2 + d)d
\chi^2 + 6 (2 + d)\chi(d + (d-1)\chi) p) n),\nonumber
\end{eqnarray}
\begin{eqnarray}
A_{30} &=& -2 d (2 + d) (2 n(n-1) (-6 \alpha_1 (2 + d) + \alpha_1 ((d-5) d-12) \chi \nonumber \\
&& + 6 \alpha_2 (1 + \chi) (2 + d  + \chi + d \chi) + \chi (12 (1 +
\chi) + d (6 + 5 \chi - d \chi))) \nonumber \\ && + (1 + d) (d \chi
(-\alpha_1 + \alpha_2 + \chi + \alpha_2 \chi) - 2 (-\alpha_1 (2 + d
+ 2 \chi) + \chi (2 + d + 2
\chi) \nonumber \\
&& + \alpha_2 (1 + \chi) (2 + d + d \chi)) p) n),\nonumber
\end{eqnarray}
\begin{eqnarray}
A_{31} &=& 2 d (2 + d) (12 n(n-1) (1 + \chi)(2 + d) \chi - (1 + d)
(-d \chi^2  + 2 (\chi (2 + d + 2 \chi)) p) n),\nonumber
\end{eqnarray}
\begin{eqnarray}
B_{30} &=& 2 (2 + d) (2 n(n-1) (-6 (\alpha_1 - \alpha_2) d (2 + d) +
(6 (2 + d) (d + \alpha_2 (2 d-1)) \nonumber \\ && + \alpha_1 (12 + d
(-12 + (d-5) d))) \chi  + (-12 + d (12 - (d-5) d) \nonumber \\ && +
\alpha_2 (d-6 + 7 d62)) \chi^2) + (1 + d) (d \chi (-\alpha_1 d + d
\chi + \alpha_2 (d + 2 (d-1) \chi)) \nonumber
\\ && - 2 (-(\alpha_1 - \alpha_2) d (2
+ d) + (-2 \alpha_1 (d-1) + d (2 + d) + 2 \alpha_2 (d-1 + d^2)) \chi
\nonumber
\\ && + (d-1) (2 + \alpha_2 d) \chi^2) p) n),  \nonumber
\end{eqnarray}
\begin{eqnarray}
B_{31} &=& 2 (2 + d) (12 n(n-1) (- \chi (d (2 + d) (1 +
\chi)-2\chi)) \nonumber \\ && \hspace{-5mm} + (1 + d) (-d^2 \chi^2 +
2 (\chi (d (2 + d) + 2 (d-1) \chi)) p) n),\nonumber
\end{eqnarray}

\begin{eqnarray}
A_{40} &=& 4 n(n-1) d (2 + d) (\alpha_2 (4 + d) (1 + \chi) (2 + d +
\chi + d \chi) \nonumber \\ && - \alpha_1 (8 (1 + \chi) + d (6 + d +
5 \chi)) + \chi (8 (1 + \chi) + d (6 + d + 5 \chi))),\nonumber
\end{eqnarray}
\begin{eqnarray}
A_{41} &=& - 4 n(n-1) d (2 + d) \chi (8 (1 + \chi) + d (6 + d + 5
\chi)),\nonumber
\end{eqnarray}
\begin{eqnarray}
B_{40} &=& 4 n(n-1) (2 + d) ((\alpha_1 - \alpha_2) d (2 + d) (4 + d)
- (2 + d) (\alpha_1 (4 - 5 d) \nonumber \\ && + (4 + d) (d +
\alpha_2 (2 d-1))) \chi - ((2 + d) (5 d-4) + \alpha_2 (-4 + d^2 (5 +
d))) \chi^2),\nonumber
\end{eqnarray}
\begin{eqnarray}
B_{41} &=& 4 n(n-1) (2 + d) (d(2 + d)(4 + d) \chi + (-8 + 6 d + 5
d^2) \chi^2),\nonumber
\end{eqnarray}

\end{document}